\documentclass[preprint]{aastex}

\newcommand \kms{km s$^{-1}$}
\newcommand \zabs{z$_{\rm abs}$}
\newcommand \zem{z$_{\rm em}$}

\usepackage{longtable}
\usepackage{graphicx}
\usepackage{natbib}
\usepackage{lscape}
\usepackage{rotating}
\usepackage{amssymb}
\usepackage{natbib}
\usepackage[T1]{fontenc}

\newcommand{\ts}{\textsuperscript}
\def \kms{km s$^{-1}$}
\def \nodata{. . .}

\def \HI{H\textsc{i}}
\def \sion{\textsc{ii}}

\begin{document}

\title{The most metal-rich damped Lyman $\alpha$ systems at $z\gtrsim1.5$ I: The Data}

\author{Trystyn A. M. Berg}
\affil{Department of Physics and Astronomy, University of Victoria, Victoria, British Columbia, V8P 1A1, Canada.}
\email{trystynb@uvic.ca}
\author{Marcel Neeleman}
\affil{Department of Physics and Center for Astrophysics and Space Sciences, UCSD, La Jolla, CA 92093, USA.}
\author{J. Xavier Prochaska}
\affil{Department of Astronomy and Astrophysics, University of California, Santa Cruz, Santa Cruz, CA, 95064, USA.}
\author{Sara L. Ellison}
\affil{Department of Physics and Astronomy, University of Victoria, Victoria, British Columbia, V8P 1A1, Canada.}
\and
\author{Arthur M. Wolfe\footnote{Deceased}}
\affil{Department of Physics and Center for Astrophysics and Space Sciences, UCSD, La Jolla, CA 92093, USA.}

\shorttitle{Metal-rich DLAs I: The Data}
\shortauthors{Berg et al.}

\keywords{Galaxies; Astrophysical Data}

\begin{abstract}
We present HIRES observations for 31 damped Lyman $\alpha$ systems, selected on the basis of their  large metal column densities from previous, lower resolution data. The measured metal column densities for Fe, Zn, S, Si, Cr, Mn, and Ni are provided for these 31 systems. Combined with previously observed large metal column density damped Lyman $\alpha$ systems, we present a sample of 44 damped Lyman $\alpha$ systems observed with high resolution spectrographs  (R$\sim30000$). These damped Lyman $\alpha$ systems probe the most chemically evolved systems at redshifts greater than 1.5. We discuss the context of our sample with the general damped Lyman $\alpha$ population, demonstrating that we are probing the top 10\% of metal column densities with our sample. In a companion paper, we will present an analysis of the sample's elemental abundances in the context of galactic chemical enrichment.

\end{abstract}

\maketitle

\section{Introduction}

Damped Lyman $\alpha$ systems (DLAs) are quasar absorption line systems with the largest neutral hydrogen column densities \citep[logN(\HI{}) $\geq$ 20.3;][]{Wolfe86} and are particularly useful to study the evolution of galaxies from redshifts $z\sim$0--5. These sight-lines through gas-rich galaxies \citep{Wolfe95} contain information on the kinematics \citep[][]{Prochaska97,Haehnelt98,Ledoux06,Neeleman13, Christensen14}, chemistry \citep[][]{Pettini90,Lu96,Prochaska01I,DZavadsky04,DZavadsky06,Ellison10,Penprase10,Ellison11,Battisti12}, and physical conditions \citep[][]{Ledoux03,Srianand05,Srianand08,York07,Milutinovic10,Tumlinson10,Ellison12,Fynbo13,Krogager13,Kanekar14} of the constituent interstellar gas. Coupled with high resolution spectrographs, observations of DLAs can provide accurate chemical compositions that allow us to understand what processes are taking place within the absorbing galaxies \citep[e.g.][]{Pettini94,Pettini97,Pettini00,Ellison01,Ledoux02,Prochaska02II,Lopez03,Akerman05}.

Previous work has demonstrated that DLAs span a wide range in metallicity where the majority of DLAs have metallicities similar to those seen in halo stars and metal-poor disk stars \citep[e.g.][]{Pettini97,Rafelski12}. Recent work by \cite{Penprase10,Cooke11,Cooke14} has focussed on the extremely metal-poor end of the distribution to study the chemical enrichment of Population III stars and metal-poor dwarf galaxies. However, little work has examined the \emph{most massive, chemically evolved} DLAs. Such studies would provide insight into dust formation \citep[][]{Pettini97,Ledoux02,Vladilo11} and nucleosynthetic constraints \citep[][]{Ellison01,Zafar14N,Zafar14Ar} within the first few Gyr of galaxy evolution. In \cite{Berg13} we presented results on the first systematic search for boron in DLAs with large metal contents, as an example of the potential for exotic element studies at high redshifts in DLAs.

A class of DLAs with large metal column densities are known as metal-strong DLAs (MSDLAs). Inspired by the first detections of exotic elements (e.g.~boron, chlorine, and germanium) in the sight-line towards the quasar FJ0812+3208 \citep{Prochaska03}, \citet[HF06]{HerbertFort06} defined a classification scheme identifying MSDLA candidates within the Sloan Digital Sky Survey (SDSS) DLA catalogues \citep{Prochaska04DR1,Prochaska05DR3} based on the strength of the Si\sion{} and Zn\sion{}  absorption lines. According to their scheme, MSDLAs require metal column densities of logN(Zn\sion) $\geq$ 13.15 \emph{or} logN(Si\sion) $ \geq$ 15.95. These limits in zinc and silicon column densities were subjectively chosen such that weak lines from rarely detected elements (such as boron, tin, and lead) could be observed in a typical high resolution spectrum purely due to the higher number of metal atoms along the sight-line.  HF06 used follow-up ESI observations to evaluate the ability to correctly identify MSDLAs from SDSS equivalent widths and visual identification, while \cite{Kaplan10} obtained N(\HI{}) for a handful of systems to determine their metallicity. However, a large collection of MSDLAs does not currently exist and no follow-up work on MSDLAs has been published since \cite{Kaplan10}. 

In this series of papers, we are interested in the metal enrichment of the most massive, metal-rich, star forming DLAs at $z\sim2$. These DLAs probe the nucleosynthetic environments similar to the metal-rich disk stars \citep[in particular the thin disk which spans from $-0.8$ $\leq$ {[Fe/H]} $\leq$ 0.2;][]{Edvardsson93}, a regime that exceeds the metallicity of the typical DLA observed. In addition, because exotic elements (e.g.~boron) have weak oscillator strengths and small abundances, the chemical evolution of these elements can also be studied in MSDLAs. In this paper, we present new data on the follow-up 31 candidate MSDLAs from the HF06 catalogue, adding to the 13 systems previously studies in the literature \citep[HF06;][]{Kaplan10}. We include a description of the high resolution spectroscopic observations and details of the abundance analysis used in \cite{Berg13} and Berg et al. (in preparation; hereafter Paper II). In addition, we revisit the MSDLA criteria defined by HF06, and discuss its significance to our sample and its impact on studying chemical evolution at high redshifts. In Paper II we will present a full analysis of the our sample's abundances in the context of local stellar populations to provide insights into the nucleosynthesis of galaxies in the first few Gyr of the universe.

\section{cMSDLA Sample}

\subsection{Sample Selection}
We have observed 31 DLAs, which were pre-selected by their \emph{very strong} metal lines in SDSS spectra (as classified by HF06). Each of these 31 DLAs were targeted for one of three specific science cases: 1) to study the chemical enrichment of DLAs \citep{Prochaska03ApJ595}, 2) for the measurement of C\textsc{i} \citep{Jorgenson10}, or 3) the detection of [C\sion{}] $\lambda$ 158 micron emission with ALMA. In all of these cases we require a relatively bright background quasar (R $\leq$ 19). The last case also requires the MSDLA candidate to fall within the observing range of ALMA (i.e. $\delta \leq 15^\circ$), and the 158 micron line be shifted into an observing band of ALMA (i.e. 1.70 $\leq$ \zabs{} $\leq$ 2.04). To increase the sample size, we supplemented this sample of DLAs with 13 additional high metal column density systems from the literature \citep[HF06;][]{Kaplan10}. These 13 DLAs were selected solely based on their high-metal content. As the total sample of 44 DLAs has been pre-selected on metal content such that there is a greater likelihood they meet the MSDLA criterion (HF06), we refer to this sample as the candidate MSDLA (hereafter referred to as the \emph{cMSDLA sample}), which will provide an excellent sample of DLAs to study the chemistry of metal enriched environments at high redshift.

\subsection{Observation Details}
The new 31 DLAs were observed with the HIgh REsolution Spectrograph \citep[HIRES;][]{Vogt94} on the Keck I telescope, spanning over several observing runs from 2005 to 2012. The 3-chip mosaic of MIT-LL 2048x4096 CCDs was used with either a 0.86'' or 1.15'' slit, resulting in a maximum full width at half maximum resolution of 6 and 8 km s$^{-1}$ respectively. All of the spectra were binned by two pixels in the spatial direction. However, for some observations the data were also binned by two in the spectral direction to reduce read noise per resolution element. This resulted in a pixel size of 2.8 km s$^{-1}$ instead of 1.4 km s$^{-1}$ for these observations. These high resolution spectra are necessary to resolve the entire kinematic structure of the metal absorption lines to derive accurate column densities and test whether the metal lines suffer from contamination or saturation. The sample of 31 new DLAs were observed over a span of 14 nights under a variety of different conditions. A journal of the observations is presented in Table \ref{tab:ObsDetails}.

\begin{table}
\scriptsize
\begin{center}
\caption{QSO targets and observation details}
\label{tab:ObsDetails}
\begin{tabular}{lccccccccc}
\hline
QSO& R.A.& Dec.& \zem{}& Magnitude& Decker& Binning& Observation& Exposure& S/N\\
 &  &  &  & (R)&  &  & date& time (s)& pixel$^{-1}$\\
\hline
J0008$-$0958& 00:08:15.3& $-$09:58:54.0& 1.95& 18.3& C1& 2x1& 2010 September 02& 15029& 11 -- 21\\
J0044+0018& 00:44:39.3& +00:18:22.7& 1.87& 18.4& C1& 2x1& 2012 January 16& 3600& 2 -- 12\\
J0058+0115& 00:58:14.3& +01:15:30.2& 2.50& 17.4& C1& 2x1& 2005 October 26& 14400& 11 -- 31\\
J0211+1241& 02:11:29.16& +12:41:10.8& 2.95& 18.9& C1& 2x1& 2011 January 25& 10200& 3 -- 10\\
J0233+0103& 02:33:33.2& +01:03:33.1& 2.06& 18.5& C5& 2x1& 2012 January 16& 3600& 2 -- 9\\
J0815+1037& 08:15:19.0& +10:37:11.5& 2.02& 18.3& C1& 2x1& 2012 January 16& 2237& 2 -- 3\\
J0927+1543& 09:27:59.8& +15:43:21.8& 1.80& 18.8& C1& 2x1& 2011 January 25& 5600& 7 -- 10\\
J0927+5823& 09:27:08.8& +58:23:19.4& 1.91& 18.3& C1& 2x1& 2011 January 26& 21600& 11 -- 20\\
J0958+0145& 09:58:22.2& +01:45:24.2& 1.96& 17.9& C1& 2x1& 2012 January 16& 3600& 2 -- 16\\
J1013+5615& 10:13:36.4& +56:15:36.4& 3.61& 18.4& C1& 2x1& 2006 January 05& 3600& 8 -- 12\\
J1024+0600& 10:24:10.4& +06:00:13.8& 2.13& 18.7& C1& 2x2& 2012 April 15& 1800& 2 -- 13\\
J1042+0628& 10:42:13.5& +06:28:53.0& 2.04& 18.8& C1& 2x2& 2012 April 15& 2400& 6 -- 12\\
J1049$-$0110& 10:49:15.4& $-$01:10:38.1& 2.12& 17.8& C5& 2x2& 2006 January 04& 4800& 10 -- 25\\
J1056+1208& 10:56:48.7& +12:08:26.8& 1.92& 17.9& C1& 2x1& 2011 January 25& 21300& 10 -- 24\\
J1106+1044& 11:06:21.4& +10:44:32.6& 1.86& 19.0& C1& 2x2& 2012 April 15& 2700& 7 -- 11\\
J1142+0701& 11:42:44.9& +07:01:03.2& 1.87& 18.7& C1& 2x1& 2012 January 16& 10200& 4 -- 8\\
J1155+0530& 11:55:38.60& +05:30:50.6& 3.48& 18.1& C1& 2x1& 2005 April 14& 7200& 12 -- 28\\
J1305+0924& 13:05:42.8& +09:24:27.8& 2.06& 18.6& C1& 2x2& 2012 April 15& 2400& 5 -- 12\\
J1310+5424& 13:10:40.24& +54:24:49.6& 1.93& 18.5& C1& 2x2& 2005 March 17& 10800& 9 -- 24\\
J1313+1441& 13:31:41.2& +14:41:40.6& 1.88& 18.2& C1& 2x1& 2006 June 03& 3600& 10 -- 15\\
J1335+0824& 13:35:32.7& +08:24:04.3& 1.91& 19.0& C1& 2x2& 2012 April 15& 3000& 5 -- 8\\
J1417+4132& 14:17:19.2& +41:32:37.0& 2.02& 18.4& C5& 2x2& 2006 June 03& 25200& 13 -- 52\\
J1454+0941& 14:54:35.2& +09:41:00.1& 1.95& 18.6& C1& 2x2& 2012 April 15& 2400& 13 -- 17\\
J1509+1113& 15:09:32.1& +11:13:13.7& 2.11& 19.0& C1& 2x2& 2012 April 15& 5194& 5 -- 12\\
J1524+1030& 15:24:30.05& +10:30:32.0& 2.06& 18.2& C1& 2x1& 2011 July 04& 9000& 4 -- 11\\
J1552+4910& 15:52:33.9& +49:10:08.3& 2.04& 18.0& C1& 2x1& 2005 May 03& 9000& 15 -- 25\\
J1555+4800& 15:55:56.9& +48:00:15.1& 3.30& 19.1& C5& 2x1& 2006 June 04& 21600& 10 -- 16\\
J1610+4724& 16:10:09.4& +47:24:44.5& 3.22& 18.6& C1& 2x1& 2006 August 18& 24900& 5 -- 12\\
J1629+0913& 16:29:03.0& +09:13:22.5& 1.99& 18.2& C1& 2x2& 2012 April 15& 2400& 9 -- 14\\
Q1755+578& 17:56:03.6& +57:48:48.0& 2.11& 18.6& C1& 2x1& 2006 August 20& 41200& 4 -- 16\\
J2241+1225& 22:41:45.1& +12:25:57.1& 2.63& 17.9& C1& 2x1& 2007 September 17& 7200& 5 -- 7\\
\end{tabular}
\end{center}
\end{table}

The raw data were reduced using the $\tt{HIRedux}$ routine, then extracted, coadded and continuum fit with $\tt{x\_continuum}$. These routines are all part of the publicly available $\tt{XIDL}$\footnote{http://www.ucolick.org/$\sim$xavier/IDL/} reduction package developed by J.X. Prochaska. Also shown in Table \ref{tab:ObsDetails} is the typical range in the signal-to-noise ratio per pixel (S/N) for each of the observed DLAs redwards of the Ly-$\alpha$ absorption.

For the majority of the cMSDLAs in Table \ref{tab:ObsDetails}, \HI{} column densities have been previously obtained from other observations with instruments such as the Echellette and Imaging Spectrometer \citep[ESI;][]{Sheinis02} on the Keck II 10m telescope and the spectra from SDSS. Unfortunately for some cases, the Ly-$\alpha$ absorption line for absorbers with redshifts less than 2.2 fall below the wavelength coverage of either of these instruments, prompting observations of DLAs with redshifts below 2.2 using the blue channel spectrograph (BCS) on the MMT telescope \citep{Kaplan10}. Even so, for 18 of the DLAs, no spectra with Ly-$\alpha$ coverage exist and we have to rely on direct measurements from the HIRES spectra. However the Ly-$\alpha$ line of a high column density absorber, such as a DLA, spreads over several orders of the HIRES spectrograph and therefore the placement of the continuum over these orders is not well-constrained. For this reason, and the difficulty of fluxing HIRES spectra \citep{Suzuki03}, we give preference to N(\HI{}) estimates made from ESI or BCS spectra where possible.

To test our determination of the \HI{} column density for the cases where we were required to use the HIRES spectra (Section 3.1), we have followed-up 4 of the absorbers using the Kast spectrograph on the Shane 3m telescope. For these observations, we used the 830 line mm$^{-1}$ grism centered at $\sim$3850 {\AA} with a 2" slit.  This resulted in a resolution of 0.63 {\AA} per pixel. The journal of observations, the average S/N of the 200 {\AA} centered around the Lyman-$\alpha$ line\footnote{Around the Ly-$\alpha$ line, Ly$\alpha$ forest absorption (which may not be resolved in low resolution observations) contaminates the spectra which reduces the S/N. The S/N measurements should be taken as a lower limit, where values of S/N $>$ 4 are acceptable.}, and the derived \HI{} column density (see Section \ref{sec:HI}) of the absorber are shown in Table \ref{tab:HIObs}. The data were reduced using the publicly available $\tt{Low}$-$\tt{Redux}$ pipeline developed by J. Hennawi, S. Burles, D. Schlegel, and J. X. Prochaska\footnote{http://www.ucolick.org/$\sim$xavier/LowRedux/}.

\begin{table}
\centering
\caption{Kast Journal of Observations}
\label{tab:HIObs}
\begin{tabular}{lccccccc}
\hline
QSO & RA & DEC & $z_{\rm{em}}$ & Date Observed & Exposure Time & S/N & logN(\HI{})\\
 & (J2000.0) & (J2000.0) & & & (s) & pixel$^{-1}$\\
\hline
J0233$+$0103 & 02 33 33.2 & $+$01 03 33.0 & 2.060 & 2011 Aug 28 & 3600 & 5 & $20.45\pm0.15$ \\
J0958$+$0145 & 09 58 22.2 & $+$01 45 24.2 & 1.960 & 2012 Feb 21 & 3600 & 4 & $20.30\pm0.15$ \\
J1313$+$1441 & 13 13 41.2 & $+$14 41 40.6 & 1.884 & 2012 Feb 20 & 3600 & 9 & $21.20\pm0.15$ \\
J1629$+$0913 & 16 29 02.9 & $+$09 13 22.5 & 1.986 & 2011 Aug 28 & 3600 & 5 & $20.80\pm0.15$ \\
\hline
\end{tabular}
\end{table}

\section{The Data}

\subsection{\HI{} Column densities}
\label{sec:HI}
We determine the \HI{} column density, N(\HI{}), of an absorber in a quasar spectrum by simultaneously fitting the continuum of the background quasar and fitting a Voigt profile to the Ly-$\alpha$ line of the absorber. This method yields an accurate \HI{} column density, if the continuum of the quasar can be accurately placed. For cases where we did not obtain follow-up observations with a blue-sensitive spectrograph, we adopt the following procedure to measure the \HI{} column density directly from the HIRES spectra. First, we observe a spectroscopic standard star of known flux at the same instrument settings as the absorber. Dividing the observed spectra of the star by the known flux yields a response function for each order. This response function is then used to find the relative flux of the absorber spectrum. \cite{Suzuki03} noted that because of the peculiarities of the HIRES instrument, they found that the fluxed spectra can be off by as much as 10\%. To account for this systematic error, we apply a 10\% adjustment to the spectra and then refit the absorber. Any differences in column densities between the measurements are then included in the uncertainty of the \HI{} column density. Figure 1 shows the fit of the Ly-$\alpha$ line for the 18 absorbers below $z\sim$ 2.2 without follow-up observations using the BCS. Note that the flux has been normalized and binned into bins of $\sim$ 20 km s$^{-1}$ for visual presentation.

\begin{figure*}
\begin{center}

\includegraphics[width=\textwidth]{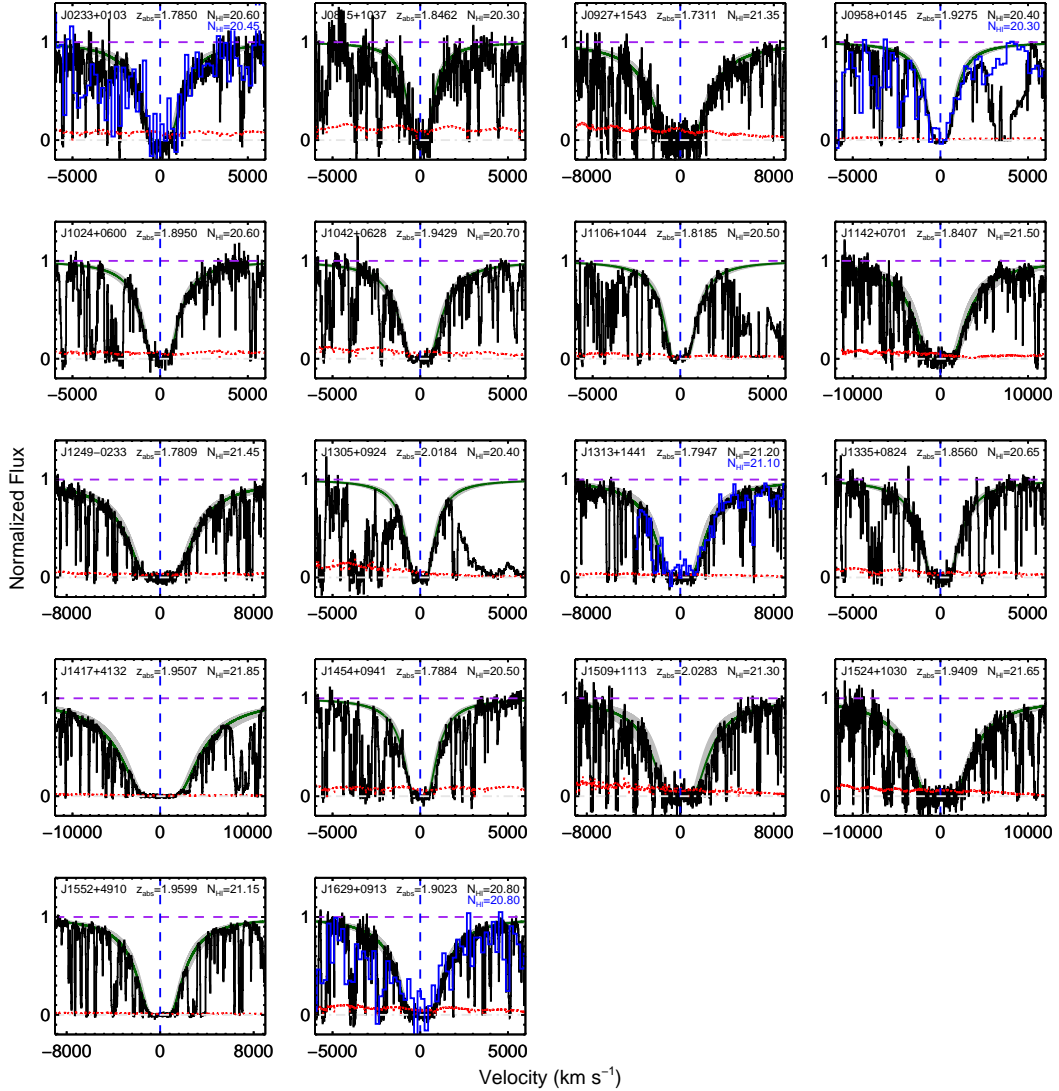}
\caption{Voigt profile fits of the Ly-$\alpha$ transition of the 18 absorbers below z $\sim$ 2.2 not observed with the BCS on MMT \citep{Kaplan10}. The gray shaded area marks the 95 \% confidence limits to the fit. The dotted (red) line is the 1-$\sigma$ error on the observation, the vertical dashed (blue) line marks the absorber's redshift measured from the metal lines, and the horizontal dashed (purple) line marks the normalized continuum of the spectrum. For the 4 DLAs observed with the Kast spectrograph, the Kast spectrum is overplotted in blue. The \HI{} column densities for these 4 DLAs observed with Kast are provided in blue below the HIRES-derived N(\HI{}).}
\label{fig:HIs}
\end{center}
\end{figure*}

To test the accuracy of using HIRES to determine N(\HI{}), we have followed up 4 absorbers with the blue-sensitive Kast spectrograph on the Shane-3m telescope (Figure \ref{fig:HIs}). The normalized spectra are shown in blue for these 4 absorbers. The resulting \HI{} column densities measured from these spectra (listed in Table \ref{tab:HIObs}) are within 0.1 dex from those measured using the HIRES spectra, well within the uncertainty of each measurement. As a result we are confident that the N(\HI{}) measurements for the remaining DLAs only observed with HIRES are accurate, and thus adopt the HIRES-derived \HI{} column densities for these 18 DLAs. The complete list of \HI{} measurements are tabulated in Table \ref{tab:HIs}.

\begin{table}
\begin{center}
\caption{\HI{} column densities of MSDLA candidates}
\label{tab:HIs}
\begin{tabular}{lcccc}
\hline
QSO& \zem{}& \zabs{}& logN(\HI{})& Reference\\
\hline
J0008$-$0958& 1.95& 1.7675& $20.85\pm0.15$& 1, 2\\
J0044+0018& 1.87& 1.7250& $20.35\pm0.10$& 2, 3\\
J0058+0115& 2.50& 2.0095& $21.10\pm0.10$& 1, 4\\
J0211+1241& 2.95& 2.5951& $20.60\pm0.15$& 2\\
J0233+0103& 2.06& 1.7850& $20.60\pm0.15$& 2\\
J0815+1037& 2.02& 1.8462& $20.30\pm0.15$& 2\\
J0927+1543& 1.80& 1.7311& $21.35\pm0.15$& 2\\
J0927+5823& 1.91& 1.6352& $20.40\pm0.15$& 3\\
J0958+0145& 1.96& 1.9275& $20.40\pm0.10$& 2\\
J1013+5615& 3.61& 2.2831& $20.70\pm0.15$& 2\\
J1024+0600& 2.13& 1.8950& $20.60\pm0.15$& 2\\
J1042+0628& 2.04& 1.9429& $20.70\pm0.15$& 2\\
J1049$-$0110& 2.12& 1.6577& $21.35\pm0.15$& 1, 3\\
J1056+1208& 1.92& 1.6093& $21.45\pm0.15$& 2, 3, 5\\
J1106+1044& 1.86& 1.8185& $20.50\pm0.15$& 2\\
J1142+0701& 1.87& 1.8407& $21.50\pm0.15$& 2\\
J1155+0530& 3.48& 3.3260& $21.05\pm0.10$& 6\\
J1305+0924& 2.06& 2.0184& $20.40\pm0.15$& 2\\
J1310+5424& 1.93& 1.8006& $21.45\pm0.15$& 2, 3, 5\\
J1313+1441& 1.88& 1.7947& $21.20\pm0.15$& 2\\
J1335+0824& 1.91& 1.8560& $20.65\pm0.15$& 2\\
J1417+4132& 2.02& 1.9509& $21.85\pm0.15$& 2, 4\\
J1454+0941& 1.95& 1.7884& $20.50\pm0.15$& 2\\
J1509+1113& 2.11& 2.0283& $21.30\pm0.15$& 2\\
J1524+1030& 2.06& 1.9409& $21.65\pm0.15$& 2\\
J1552+4910& 2.04& 1.9599& $21.15\pm0.15$& 2\\
J1555+4800& 3.30& 2.3911& $21.50\pm0.15$& 2\\
J1610+4724& 3.22& 2.5066& $21.00\pm0.15$& 3\\
J1629+0913& 1.99& 1.9023& $20.80\pm0.10$& 2\\
Q1755+578& 2.11& 1.9692& $21.40\pm0.15$& 2\\
J2241+1225& 2.63& 2.4179& $21.15\pm0.10$& 2\\
\end{tabular}

\textsc{References}--
	(1) \cite{HerbertFort06}.
	(2) This~Work.
	(3) \cite{Kaplan10}.
	(4) \cite{Berg13}.
	(5) \cite{Prochaska08}.
	(6) \cite{Wolfe08}.
\end{center}
\end{table}

\subsection{Metal Column Densities}

As in previous, large surveys of metal column densities \citep[e.g.][]{Prochaska01I}, all metal column densities measured for the cMSDLA sample (given in Table \ref{tab:ColSumm}) were obtained using the apparent optical depth method (AODM) outlined by \cite{Savage91}. The AODM provides accurate column densities for non-saturated and non-blended lines, and is faster than fitting Voigt profiles for each individual system. The AODM sums the optical depth ($\tau$) of an unsaturated absorption line (at wavelength $\lambda$, with oscillator strength $f$) and is converted to a column density ($N$) using \begin{equation}
 N=\frac{m_{e}c}{\pi e^{2} f \lambda}\int \tau dv \label{eq:AODM}
\end{equation} where the integral of the optical depth sums over each pixel in velocity space. The limits for the optical depth integration are chosen to contain the absorption profile that is common to all non-contaminated transitions. For lines that were either blended or saturated (and no other \emph{clean} transitions were available for the same species), the derived AODM column density was taken as an upper or lower limit (respectively). The errors quoted on the column densities ($N_{err}$) were determined from the error spectrum using \begin{equation}
N_{err}=\frac{m_{e}c}{\pi e^{2} f \lambda}\left(\sum \left(\frac{I_{err}}{I_{spec}} \Delta v\right)^{2} \right)^{0.5}\label{eq:eAODM}
\end{equation} where $I_{err}$ and $I_{spec}$ are the fluxes in the error and observed spectra (respectively), and $\Delta v$ is the velocity width of the pixel. The error spectrum only accounts for photon noise, and not continuum errors. A minimum error of 0.05 dex is adopted for all metal columns to account for any systematic errors (such as continuum placement).

For each DLA sight-line, all transitions seen for the elements Fe, Zn, S, Si, Cr, Mn, and Ni  are shown in Figures 2--32\footnote{Figures 3--32 are provided in the published version of this paper in PASP (online edition only). Figure 2 is provided as an example.} , with the AODM velocity limits and column densities given in Tables \ref{tab:J00080958}--\ref{tab:J22411225}.  The adopted column densities  in Tables \ref{tab:J00080958}--\ref{tab:J22411225} (N$_{adopt}$) are determined from the included lines using a weighted mean\footnote{Each line is weighted by $N_{err}^{-2}$ (solely derived from the error spectrum) in order to reflect the quality of the spectra at each line.}. In cases where we suspect \emph{mild} saturation, we fit a Voigt profile using \textsc{VPFIT}\footnote{http://www.ast.cam.ac.uk/$\sim$rfc/vpfit.html} to test whether the AODM column densities are accurate. We discuss any subtleties of the measurements on a case by case basis below. For all plots of the metal line profiles, the horizontal dashed line shows the continuum, while the vertical dotted lines show the velocity limits for the AODM integration to obtain the column density. All bad pixels within each plot are grayed out. All atomic data were taken from \cite{Morton03}.

\subsubsection{J0008$-$0958}

 The S\sion{} $\lambda$ 1253 absorption is strongly blended with the Ly$\alpha$ forest, and is ignored from the adopted S\sion{} column density. We adopt the Zn column density obtained solely from the Zn\sion{} $\lambda$ 2026 line. There appears to be some slight blending at $\sim-160$\kms{} at the Zn\sion{} $\lambda$ 2062  line, likely from Cr\sion{} $\lambda$ 2062  contamination, causing an overestimate of the obtained column density for the redder transition.

\begin{figure}
\begin{center}
\includegraphics[width=1.0\textwidth]{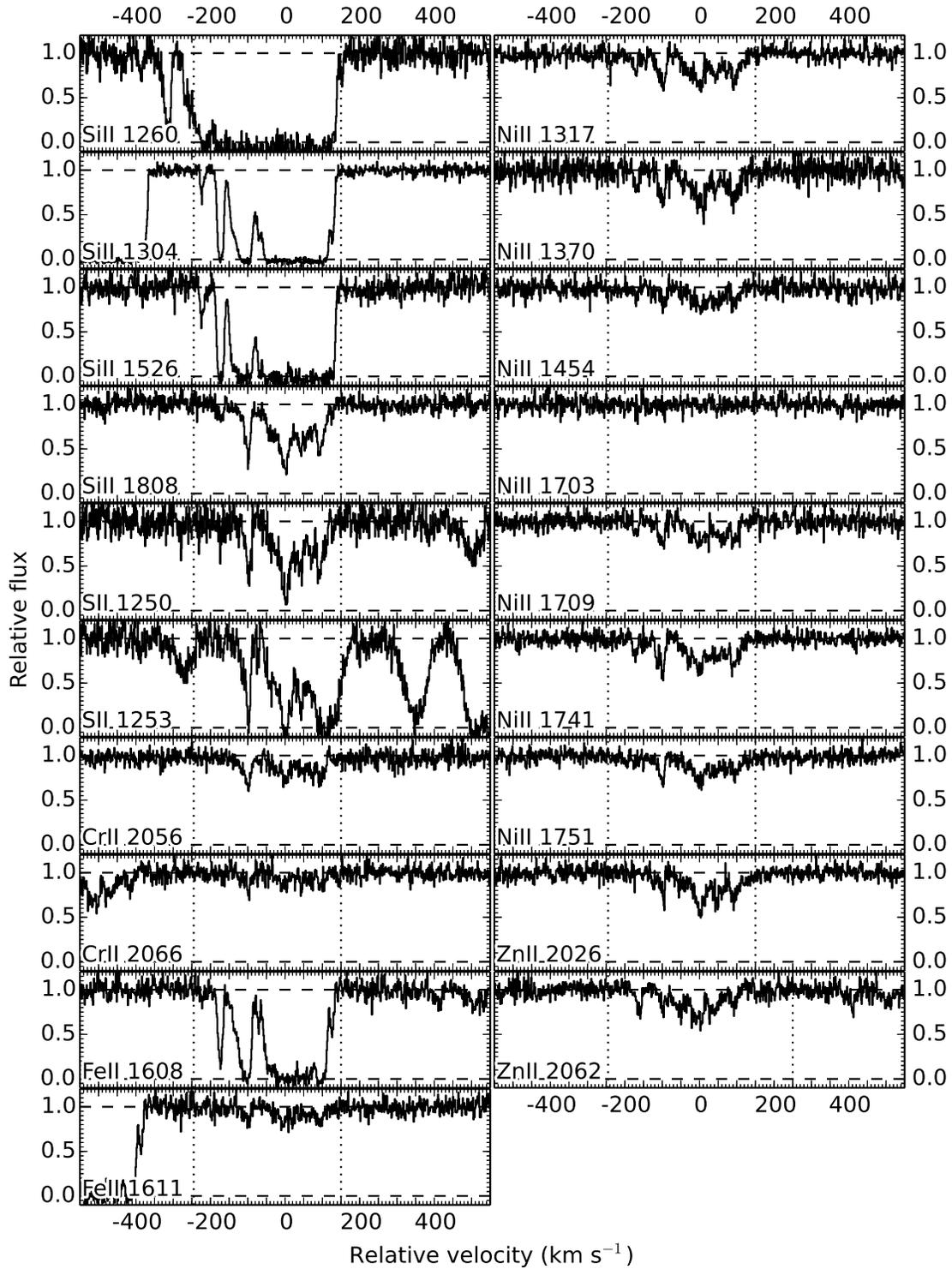}
\caption{Metal line velocity profiles for J0008$-$0958.}
\label{fig:J00080958metal}
\end{center}
\end{figure}

\subsubsection{J0044+0018}

The DLA towards J0044+0018 is believed to have a large amount of cold gas, as indicated by the detection of C\textsc{i}. As a result, the DLA likely has a large amount of Mg\textsc{i} present, contaminating the Zn\sion{} $\lambda$ 2026  line (e.g.~HF06 and references therein). It is clear in Figure 3 that the Zn\sion{} $\lambda$ 2026  line has a slight double peak for each absorption component relative to the other lines, due to blending from Mg\textsc{i}. We therefore treat the measured N(Zn\sion{}) as an upper limit.

\subsubsection{J0058+0115}

For the DLA towards J0058+0115, we quote a detection of the Zn\sion{} column density, despite having a dead pixel within the absorption at $\sim0$\kms{}. The dead pixel only contributes $\sim$0.02 dex to the total column density (within the quoted error), and is therefore adopted without any corrections.

\subsubsection{J0211+1241}

Although the S\sion{} $\lambda$ 1253 line is covered in our spectrum, there is severe contamination with a Ly$\alpha$ forest cloud, so no measurement is provided.

\subsubsection{J0233+0103}

Although both Si\sion{} $\lambda$ 1304 and 1526 lines appear saturated,  we can constrain N(Si\sion{}) in the DLA towards J0233+0103 using the upper limit of the column density derived using Si\sion{} $\lambda$ 1808  line. With both upper and lower limits obtained from these three lines, the total Si\sion{} column density should be N(Si\sion{}) $\sim14.77\pm0.05$.

\subsubsection{J0815+1037}
We note that the error on the column density derived for Si \sion{} $\lambda$ 1808 in the DLA towards J0815+1037 is very high (0.45 dex) compared to the typical error in metal column densities. The adopted error reflects the low S/N in the spectrum (S/N $\sim$ 2--3) for the weak line.

\subsubsection{J0927+1543}

For the DLA detected towards J0927+1543 sight-line,  both Fe\sion{} $\lambda$ 1608 and 1611 lines provide tight constraints on the true N(Fe\sion{}) of the system, and are averaged together to obtain the final adopted column density.

\subsubsection{J0958+0145}

Although there is clearly some absorption of Zn\sion{} $\lambda$ 2026 in the spectrum of J0958+0145 with the same expected absorption profile for a line of this strength (see Figure 10), we quote N(Zn\sion{}) as an upper limit. As the strongest component of the Zn\sion{} $\lambda$ 2026 absorption is offset from by $\sim 10$\kms{} relative to other metal lines, we are cautious to deem our measured column density as a detection. We therefore conservatively quote the adopted column density as an upper limit.

\subsubsection{J1013+5615}
In the DLA towards J1013+5615, both Cr\sion{} $\lambda$ 2056 and Zn\sion{} $\lambda$ 2026 contain bad pixels within the absorption features. Although the determined column density was derived including these bad pixels; the bad pixels have a negligible effect on the total column density (0.002 and 0.021 dex, respectively). The final column densities are not adjusted as a result of the bad pixels.

\subsubsection{J1042+0628}

The lower and upper limits in the derived column densities for Fe\sion{} $\lambda$ 1608 and 1611 (respectively) provide constraining bounds on the expected N(Fe\sion{}) in J1042+0628. As a result, we assume an average of the two limits (N(Fe\sion{})=$15.00\pm0.15$) as the adopted column density for this DLA.

\subsubsection{J1056+1208}
Even though the Si\sion{} $\lambda$ 1808 line is slightly saturated, we can get a robust measurement by using a chi-squared fitting routine, VPFIT. We tied the velocity structure of the Si\sion{} line to the unsaturated Zn\sion{} line and measured the column density. We compared the VPFIT-derived column density to the AODM-derived column density and found the two to be in agreement within the error of the measurements.

\subsubsection{J1142+0701}
Si \sion{} $\lambda$ 1808 appears saturated in the spectrum of the DLA towards J1142+0701. To verify the effects of saturation, VPFIT was used to determine the column density (similar to the procedure describe for the DLA in J1056+1208). The VPFIT column density agrees with the AODM derived column density given.

\subsubsection{J1305+0924}
For the DLA in the J1305+0924 spectrum, the absorption feature from S\sion{} $\lambda$ 1259 is slightly blended with a Ly$\alpha$ cloud in the highest velocity component (75 \kms{}). With the lack of absorption beyond 75 \kms{} seen in the other metal lines, truncating the AODM limit for S\sion{} $\lambda$ 1259 at 75 \kms{} should have a negligible effect on the total column density derived, and is still consistent with the other S\sion{} lines.

\subsubsection{J1313+1441}

For this DLA, we are only able to obtain lower limits on N(S\sion{}) from the S\sion{} $\lambda$ 1250 and 1253 lines due to saturation. Although S\sion{} $\lambda$ 1253 provides the more constraining limit, there appears to be unrelated absorption at $\sim 100$\kms{}. As a result, we ignore this more constraining limit. In \cite{Berg13}, we had originally reported the sulphur column density as a detection rather than a limit, as it is unclear whether the S\sion{} $\lambda$ 1250 line is saturated. To be conservative, we have decided to adopt the derived column density as a lower limit. Similar to the DLA towards J1056+1208; we have used VPFIT to check the derived column density from Si\sion{} $\lambda$ 1808 and find agreement with the AODM derived column density.

\subsubsection{J1417+4132}
Although a limit of  logN(S) $>$ 15.8 was initially reported from the data in \cite{Berg13} as it appears saturated, the S\sion{} data is too blended to place a constraining limit on the derived column density. To be consistent with other DLAs we have previously discussed, we do not include such a limit as we are unsure of the role of saturation and contamination to place a meaningful limit on the column density.

\subsubsection{J1555+4800}

The J1555+4800 spectrum shows a Ly$\alpha$ cloud blended with S\sion{} $\lambda$ 1253, making it impossible to determine the amount of saturation and blending to provide a meaningful limit. However, it is worth noting that assuming that the absorption is real, a lower limit of logN(S) $>$ 15.88  was measured for this system \citep{Berg13}.

\subsubsection{Q1755+578}
The DLA towards Q1755+578 is a system with a large amount of C\textsc{i} (i.e. a significant amount of cold gas). It is clear in Figure 31 for Zn\sion{} $\lambda$ 2056 there is an additional narrow absorption at $\sim50$\kms{}, that is likely due to Mg\textsc{i} absorption. From Voigt profile modelling of this complex system, we obtained a column density of N(Zn\sion{})$=13.85 \pm 0.05$.

\subsubsection{J2241+1225}

The S\sion{} column density for J2241+1225  challenging due to uncertain levels of saturation of the S\sion{} $\lambda$ 1253 and 1259. As S\sion{} $\lambda$ 1259 appears to have a slight excess of absorption at $\sim-60$\kms{} relative to the typical absorption profile for this system, we adopt the value from S\sion{} $\lambda$ 1253. Although it is likely that the line is not saturated \citep[as originally claimed in][N(S\sion{})$=14.94 \pm 0.05$]{Berg13}, we remain conservative and keep this value as a lower limit. 

\begin{landscape}
\begin{table}
\scriptsize
\begin{center}
\caption{Summary of Column Densities}
\label{tab:ColSumm}
\begin{tabular}{lcccccccccc}
\hline
QSO& \zem{}& \zabs{}& logN(H\textsc{i})& logN(Si\sion{})& logN(S\sion{})& logN(Mn\sion{})& logN(Cr\sion{})& logN(Fe\sion{})& logN(Ni\sion{})& logN(Zn\sion{})\\
\hline
J0008$-$0958& 1.95& 1.7675& $20.85\pm0.15$& $16.04 \pm 0.05$& $15.84 \pm 0.05$& \nodata{}& $13.91 \pm 0.05$& $15.62 \pm 0.05$& $14.46 \pm 0.05$& $13.31 \pm 0.05$\\
J0044+0018& 1.87& 1.7250& $20.35\pm0.10$& $15.34 \pm 0.05$& $15.27 \pm 0.05$& \nodata{}& $< 13.04$& $> 14.77$& $13.89 \pm 0.05$& $< 12.61$\\
J0058+0115& 2.50& 2.0095& $21.10\pm0.10$& $> 15.55$& $15.40 \pm 0.05$& \nodata{}& $13.54 \pm 0.05$& $15.18 \pm 0.05$& $14.16 \pm 0.05$& $12.95 \pm 0.05$\\
Q0201+36& 2.91& 2.4628& $20.38\pm0.04$& $15.53 \pm 0.05$& $15.29 \pm 0.05$& \nodata{}& $13.24 \pm 0.05$& $15.01 \pm 0.05$& $14.03 \pm 0.05$& $12.76 \pm 0.05$\\
J0211+1241& 2.95& 2.5951& $20.60\pm0.15$& $15.53 \pm 0.08$& \nodata{}& \nodata{}& \nodata{}& $15.06 \pm 0.05$& $14.07 \pm 0.05$& \nodata{}\\
J0233+0103& 2.06& 1.7850& $20.60\pm0.15$& $14.77 \pm 0.05$& \nodata{}& \nodata{}& \nodata{}& $14.62 \pm 0.05$& $13.61 \pm 0.11$& \nodata{}\\
Q0458$-$02& 2.29& 2.0396& $21.65\pm0.09$& $16.04 \pm 0.05$& \nodata{}& \nodata{}& $13.76 \pm 0.05$& $15.38 \pm 0.05$& $14.18 \pm 0.05$& $13.13 \pm 0.05$\\
FJ0812+3208& 2.70& 2.6263& $21.35\pm0.10$& $15.98 \pm 0.05$& $15.63 \pm 0.07$& $< 13.00$& $13.36 \pm 0.05$& $15.09 \pm 0.05$& $13.89 \pm 0.05$& $13.15 \pm 0.05$\\
J0815+1037& 2.02& 1.8462& $20.30\pm0.15$& $15.38 \pm 0.45$& \nodata{}& \nodata{}& \nodata{}& $> 14.87$& $13.74 \pm 0.12$& \nodata{}\\
J0927+1543& 1.80& 1.7311& $21.35\pm0.15$& $15.99 \pm 0.05$& \nodata{}& \nodata{}& $13.83 \pm 0.05$& $15.14 \pm 0.24$& $14.17 \pm 0.05$& $13.38 \pm 0.05$\\
J0927+5823& 1.91& 1.6352& $20.40\pm0.15$& $15.72 \pm 0.05$& $15.61 \pm 0.05$& \nodata{}& \nodata{}& $> 15.27$& $14.44 \pm 0.05$& $13.29 \pm 0.05$\\
J0958+0145& 1.96& 1.9275& $20.40\pm0.10$& $14.84 \pm 0.06$& $14.44 \pm 0.05$& \nodata{}& \nodata{}& $14.23 \pm 0.05$& $13.37 \pm 0.07$& $< 12.00$\\
J1010+0003& 1.40& 1.2651& $21.52\pm0.07$& \nodata{}& \nodata{}& \nodata{}& $13.54 \pm 0.07$& $15.26 \pm 0.05$& \nodata{}& $12.96 \pm 0.06$\\
J1013+5615& 3.61& 2.2831& $20.70\pm0.15$& $16.14 \pm 0.05$& \nodata{}& \nodata{}& $13.79 \pm 0.05$& $> 15.45$& \nodata{}& $13.56 \pm 0.05$\\
J1024+0600& 2.13& 1.8950& $20.60\pm0.15$& $15.81 \pm 0.05$& $15.45 \pm 0.05$& \nodata{}& \nodata{}& $15.27 \pm 0.08$& $14.02 \pm 0.05$& \nodata{}\\
J1042+0628& 2.04& 1.9429& $20.70\pm0.15$& $15.40 \pm 0.08$& $15.08 \pm 0.05$& \nodata{}& \nodata{}& $15.00 \pm 0.15$& $< 13.78$& \nodata{}\\
J1049$-$0110& 2.12& 1.6577& $21.35\pm0.15$& $15.80 \pm 0.05$& $15.47 \pm 0.05$& \nodata{}& $13.49 \pm 0.05$& $15.17 \pm 0.05$& $14.25 \pm 0.05$& $13.14 \pm 0.05$\\
J1056+1208& 1.92& 1.6093& $21.45\pm0.15$& $16.48 \pm 0.09$& $> 16.15$& \nodata{}& $14.04 \pm 0.05$& $15.81 \pm 0.05$& $14.69 \pm 0.05$& $13.76 \pm 0.05$\\
J1106+1044& 1.86& 1.8185& $20.50\pm0.15$& $> 15.22$& $15.33 \pm 0.05$& \nodata{}& \nodata{}& $> 15.15$& $14.02 \pm 0.05$& \nodata{}\\
J1142+0701& 1.87& 1.8407& $21.50\pm0.15$& $16.15 \pm 0.13$& \nodata{}& \nodata{}& $13.70 \pm 0.05$& $15.47 \pm 0.05$& $14.01 \pm 0.05$& $13.29 \pm 0.05$\\
J1155+0530& 3.48& 3.3260& $21.05\pm0.10$& $15.94 \pm 0.05$& $15.40 \pm 0.05$& \nodata{}& $13.36 \pm 0.09$& $15.37 \pm 0.05$& $14.07 \pm 0.05$& $12.89 \pm 0.07$\\
J1159+0112& 1.99& 1.9440& $21.70\pm0.10$& $15.95 \pm 0.05$& \nodata{}& $< 13.26$& $13.82 \pm 0.05$& $15.49 \pm 0.05$& $14.20 \pm 0.05$& $13.11 \pm 0.06$\\
J1200+4015& 3.36& 3.2200& $20.85\pm0.10$& $> 15.21$& $15.36 \pm 0.05$& \nodata{}& $13.53 \pm 0.05$& $15.31 \pm 0.05$& $14.18 \pm 0.05$& $12.86 \pm 0.05$\\
J1249$-$0233& 2.12& 1.7809& $21.45\pm0.15$& $> 15.11$& $15.53 \pm 0.05$& \nodata{}& \nodata{}& \nodata{}& $14.29 \pm 0.05$& $13.15 \pm 0.05$\\
J1305+0924& 2.06& 2.0184& $20.40\pm0.15$& $15.75 \pm 0.05$& $15.39 \pm 0.05$& \nodata{}& \nodata{}& $15.21 \pm 0.14$& $14.36 \pm 0.05$& \nodata{}\\
J1310+5424& 1.93& 1.8006& $21.45\pm0.15$& $16.44 \pm 0.05$& $> 16.05$& \nodata{}& $13.99 \pm 0.05$& $15.64 \pm 0.05$& $14.45 \pm 0.05$& $13.57 \pm 0.05$\\
J1313+1441& 1.88& 1.7947& $21.20\pm0.15$& $16.12 \pm 0.05$& $> 15.75$& \nodata{}& $13.61 \pm 0.05$& $15.55 \pm 0.05$& $14.27 \pm 0.05$& $13.30 \pm 0.05$\\
J1335+0824& 1.91& 1.8560& $20.65\pm0.15$& $15.73 \pm 0.05$& $15.29 \pm 0.05$& $13.70 \pm 0.10$& $13.81 \pm 0.05$& $> 15.17$& $14.29 \pm 0.05$& \nodata{}\\
J1417+4132& 2.02& 1.9509& $21.85\pm0.15$& $> 16.42$& \nodata{}& \nodata{}& $14.04 \pm 0.05$& $15.58 \pm 0.05$& $14.55 \pm 0.05$& $13.55 \pm 0.05$\\
J1454+0941& 1.95& 1.7884& $20.50\pm0.15$& $15.47 \pm 0.05$& $15.25 \pm 0.06$& \nodata{}& $13.30 \pm 0.09$& $15.02 \pm 0.12$& $13.85 \pm 0.05$& $12.72 \pm 0.05$\\
J1509+1113& 2.11& 2.0283& $21.30\pm0.15$& $16.04 \pm 0.05$& $15.69 \pm 0.05$& \nodata{}& \nodata{}& $15.48 \pm 0.07$& $14.41 \pm 0.05$& \nodata{}\\
J1524+1030& 2.06& 1.9409& $21.65\pm0.15$& $> 16.24$& $> 15.63$& \nodata{}& $13.57 \pm 0.05$& $15.44 \pm 0.05$& $14.53 \pm 0.05$& $> 13.53$\\
J1552+4910& 2.04& 1.9599& $21.15\pm0.15$& $15.98 \pm 0.05$& $15.34 \pm 0.05$& $13.39 \pm 0.05$& $13.74 \pm 0.05$& $15.47 \pm 0.05$& $14.24 \pm 0.05$& $12.93 \pm 0.05$\\
J1555+4800& 3.30& 2.3911& $21.50\pm0.15$& $16.55 \pm 0.05$& \nodata{}& \nodata{}& $14.19 \pm 0.05$& $15.84 \pm 0.05$& $14.78 \pm 0.05$& $< 13.95$\\
J1604+3951& 3.15& 3.1633& $21.75\pm0.00$& $> 15.31$& $15.71 \pm 0.05$& \nodata{}& \nodata{}& $15.47 \pm 0.05$& $14.24 \pm 0.05$& $13.12 \pm 0.05$\\
J1610+4724& 3.22& 2.5066& $21.00\pm0.15$& $16.16 \pm 0.05$& \nodata{}& \nodata{}& $13.90 \pm 0.05$& $15.62 \pm 0.05$& $14.58 \pm 0.05$& $13.56 \pm 0.05$\\
J1629+0913& 1.99& 1.9023& $20.80\pm0.10$& $15.32 \pm 0.06$& $15.24 \pm 0.05$& \nodata{}& $< 13.21$& $> 14.93$& $13.75 \pm 0.13$& $12.68 \pm 0.08$\\
Q1755+578& 2.11& 1.9692& $21.40\pm0.15$& $16.58 \pm 0.05$& $> 16.12$& $13.83 \pm 0.05$& $14.09 \pm 0.05$& $15.79 \pm 0.05$& $14.75 \pm 0.05$& $13.85 \pm 0.05$\\
J2100$-$0641& 3.14& 3.0924& $21.05\pm0.15$& $15.88 \pm 0.05$& $15.49 \pm 0.05$& \nodata{}& $13.59 \pm 0.05$& $15.37 \pm 0.05$& $14.23 \pm 0.05$& $13.24 \pm 0.05$\\
J2222$-$0946& 2.93& 2.3543& $20.55\pm0.15$& $15.68 \pm 0.05$& $15.37 \pm 0.05$& \nodata{}& \nodata{}& $15.06 \pm 0.08$& $14.04 \pm 0.05$& \nodata{}\\
Q2230+02& 2.15& 1.8644& $20.85\pm0.08$& $15.65 \pm 0.05$& $15.29 \pm 0.05$& \nodata{}& $13.40 \pm 0.05$& $15.19 \pm 0.05$& $14.13 \pm 0.05$& $12.80 \pm 0.05$\\
J2241+1225& 2.63& 2.4179& $21.15\pm0.10$& $> 14.67$& $> 15.01$& $13.36 \pm 0.13$& \nodata{}& $15.02 \pm 0.08$& $13.83 \pm 0.05$& \nodata{}\\
J2340$-$0053& 2.09& 2.0545& $20.35\pm0.15$& $15.23 \pm 0.05$& $14.95 \pm 0.05$& \nodata{}& \nodata{}& $14.98 \pm 0.05$& $13.81 \pm 0.05$& $12.63 \pm 0.07$\\
Q2342+34& 2.92& 2.9082& $21.10\pm0.10$& $15.62 \pm 0.05$& $15.19 \pm 0.05$& \nodata{}& $13.23 \pm 0.11$& $14.91 \pm 0.07$& $13.81 \pm 0.05$& $< 12.60$\\
\end{tabular}
\end{center}
\end{table}
\end{landscape}

\setcounter{figure}{32}

\section{Summary}
\subsection{Sample Properties}

A summary of all the properties of the entire cMSDLA sample (literature and new observations) is provided in Table \ref{tab:DLAsum}. Included in Table \ref{tab:DLAsum} are the metallicities \citep[following the scheme in][R12]{Rafelski12}, whether or not the DLA is a bona fide MSDLA, and the velocity width of the inner 90\% of the metal lines \citep[$\Delta v_{90}$; see][]{Prochaska97}. It is interesting to note that the majority of DLAs in our sample have $\Delta v_{90}$ $\gtrsim$ $100$\kms{}, implying that these systems live in some of the most massive dark matter halos in which DLAs reside \citep[see][]{Neeleman13}.

\begin{table}
\scriptsize
\begin{center}
\caption{Summary of cMSDLA Sample}
\label{tab:DLAsum}
\begin{tabular}{lcccccc}
\hline
QSO& \zabs{}& logN(\HI{})& [M/H] (elem)& MSDLA?& $\Delta v_{90}$ (line)& Ref.\\
 &  &  &  &  &  km s$^{-1}$&  \\
\hline
J0008$-$0958& $1.7675$& $20.85\pm0.15$& $-0.16\pm0.16$ (S)& True& 216 (Si\sion{} 1808)& 1,2\\
J0044+0018& $1.7250$& $20.35\pm0.10$& $-0.23\pm0.11$ (S)& False& 172 (S\sion{} 1253)& 2\\
J0058+0115& $2.0095$& $21.10\pm0.10$& $-0.85\pm0.11$ (S)& False& 195 (S\sion{} 1253)& 1,2\\
Q0201+36& $2.4628$& $20.38\pm0.04$& $-0.24\pm0.07$ (S)& False& 200 (Si\sion{} 1808)& 1,3,4,5,6\\
J0211+1241& $2.5951$& $20.60\pm0.15$& $-0.58\pm0.17$ (Si)& False& 48 (Ni\sion{} 1741)& 2\\
J0233+0103& $1.7850$& $20.60\pm0.15$& $-1.34\pm0.16$ (Si)& False& 97 (Fe\sion{} 1608)& 2\\
Q0458$-$02& $2.0396$& $21.65\pm0.09$& $-1.12\pm0.10$ (Si)& True& 84 (Cr\sion{} 2056)& 1,2,5,7,8\\
FJ0812+3208& $2.6263$& $21.35\pm0.10$& $-0.87\pm0.12$ (S)& True& 56 (S\sion{} 1250)& 1,2,9,10\\
J0815+1037& $1.8462$& $20.30\pm0.15$& $-0.43\pm0.47$ (Si)& False& -99 (no line)& 2\\
J0927+1543& $1.7311$& $21.35\pm0.15$& $-0.87\pm0.16$ (Si)& True& 220 (Si\sion{} 1808)& 1,2\\
J0927+5823& $1.6352$& $20.40\pm0.15$& $0.06\pm0.16$ (S)& True& 208 (Si\sion{} 1808)& 1,2\\
J0958+0145& $1.9275$& $20.40\pm0.10$& $-1.11\pm0.11$ (S)& False& 56 (S\sion{} 1259)& 1,2\\
J1010+0003& $1.2651$& $21.52\pm0.07$& $-1.19\pm0.10$ (Zn)& False& 36 (Ni\sion{} 1709)& 1,2,11,12\\
J1013+5615& $2.2831$& $20.70\pm0.15$& $-0.07\pm0.16$ (Si)& True& 213 (Si\sion{} 1808)& 1,2\\
J1024+0600& $1.8950$& $20.60\pm0.15$& $-0.30\pm0.16$ (S)& False& 161 (Si\sion{} 1808)& 2\\
J1042+0628& $1.9429$& $20.70\pm0.15$& $-0.77\pm0.16$ (S)& False& 135 (S\sion{} 1253)& 2\\
J1049$-$0110& $1.6577$& $21.35\pm0.15$& $-1.03\pm0.16$ (S)& False& 330 (Si\sion{} 1808)& 1,2\\
J1056+1208& $1.6093$& $21.45\pm0.15$& $-0.48\pm0.18$ (Si)& True& 124 (Ni\sion{} 1370)& 1,2\\
J1106+1044& $1.8185$& $20.50\pm0.15$& $-0.32\pm0.16$ (S)& False& 203 (S\sion{} 1253)& 2\\
J1142+0701& $1.8407$& $21.50\pm0.15$& $-0.86\pm0.20$ (Si)& True& 52 (Ni\sion{} 1370)& 2\\
J1155+0530& $3.3260$& $21.05\pm0.10$& $-0.80\pm0.11$ (S)& False& 220 (S\sion{} 1250)& 1,2\\
J1159+0112& $1.9440$& $21.70\pm0.10$& $-1.26\pm0.11$ (Si)& True& 84 (Ni\sion{} 1741)& 1,2,13,14,15\\
J1200+4015& $3.2200$& $20.85\pm0.10$& $-0.64\pm0.11$ (S)& False& 127 (Ni\sion{} 1317)& 1,2,16\\
J1249$-$0233& $1.7809$& $21.45\pm0.15$& $-1.07\pm0.16$ (S)& True& 152 (S\sion{} 1250)& 1,2,17\\
J1305+0924& $2.0184$& $20.40\pm0.15$& $-0.16\pm0.16$ (S)& False& 135 (S\sion{} 1253)& 2\\
J1310+5424& $1.8006$& $21.45\pm0.15$& $-0.52\pm0.16$ (Si)& True& 86 (Ni\sion{} 1751)& 1,2\\
J1313+1441& $1.7947$& $21.20\pm0.15$& $-0.59\pm0.16$ (Si)& True& 147 (Zn\sion{} 2026)& 1,2\\
J1335+0824& $1.8560$& $20.65\pm0.15$& $-0.51\pm0.16$ (S)& False& 166 (S\sion{} 1253)& 2\\
J1417+4132& $1.9509$& $21.85\pm0.15$& $-0.93\pm0.16$ (Zn)& True& 114 (Zn\sion{} 2026)& 1,2\\
J1454+0941& $1.7884$& $20.50\pm0.15$& $-0.40\pm0.16$ (S)& False& 81 (S\sion{} 1253)& 2\\
J1509+1113& $2.0283$& $21.30\pm0.15$& $-0.76\pm0.16$ (S)& True& 101 (S\sion{} 1253)& 2\\
J1524+1030& $1.9409$& $21.65\pm0.15$& $-1.36\pm0.16$ (Fe)& True& 201 (Ni\sion{} 1709)& 1,2\\
J1552+4910& $1.9599$& $21.15\pm0.15$& $-0.96\pm0.16$ (S)& True& 112 (Si\sion{} 1808)& 1,2\\
J1555+4800& $2.3911$& $21.50\pm0.15$& $-0.46\pm0.16$ (Si)& True& 199 (Ni\sion{} 1741)& 1,2\\
J1604+3951& $3.1633$& $21.75\pm0.00$& $-1.19\pm0.05$ (S)& False& 429 (S\sion{} 1250)& 1,2,18\\
J1610+4724& $2.5066$& $21.00\pm0.15$& $-0.35\pm0.16$ (Si)& True& 155 (Si\sion{} 1808)& 1,2,17\\
J1629+0913& $1.9023$& $20.80\pm0.10$& $-0.71\pm0.11$ (S)& False& 117 (S\sion{} 1253)& 2\\
Q1755+578& $1.9692$& $21.40\pm0.15$& $-0.33\pm0.16$ (Si)& True& 364 (Ni\sion{} 1741)& 1,2\\
J2100$-$0641& $3.0924$& $21.05\pm0.15$& $-0.71\pm0.16$ (S)& True& 187 (S\sion{} 1250)& 1,2,17\\
J2222$-$0946& $2.3543$& $20.55\pm0.15$& $-0.33\pm0.16$ (S)& False& 173 (S\sion{} 1253)& 1,2,17,19\\
Q2230+02& $1.8644$& $20.85\pm0.08$& $-0.71\pm0.10$ (S)& False& 172 (Si\sion{} 1808)& 1,2,5,7,8,14,15\\
J2241+1225& $2.4179$& $21.15\pm0.10$& $-1.28\pm0.13$ (Fe)& False& 65 (S\sion{} 1253)& 1,2\\
J2340$-$0053& $2.0545$& $20.35\pm0.15$& $-0.55\pm0.16$ (S)& False& 138 (Si\sion{} 1808)& 1,2,10\\
Q2342+34& $2.9082$& $21.10\pm0.10$& $-1.06\pm0.11$ (S)& False& 100 (Si\sion{} 1808)& 1,2,9,10\\
\hline
\end{tabular}

\textsc{References}--
	(1) \cite{Berg13}.
	(2) This Work.
	(3) \cite{DLAcat5}.
	(4) \cite{Pettini97}.
	(5) \cite{DLAcat23}.
	(6) \cite{DLAcat35}.
	(7) \cite{DLAcat13}.
	(8) \cite{DLAcat24}.
	(9) \cite{DLAcat43}.
	(10) \cite{DLAcat64}.
	(11) \cite{DLAcat60}.
	(12) \cite{DLAcat78}.
	(13) \cite{DLAcat18}.
	(14) \cite{DLAcat54}.
	(15) \cite{DLAcat66}.
	(16) \cite{Rafelski12}.
	(17) \cite{HerbertFort06}.
	(18) \cite{DLAcat80}.
	(19) \cite{DLAcat102}.

\end{center}
\end{table}

\begin{figure}
\begin{center}
\includegraphics[width=\textwidth]{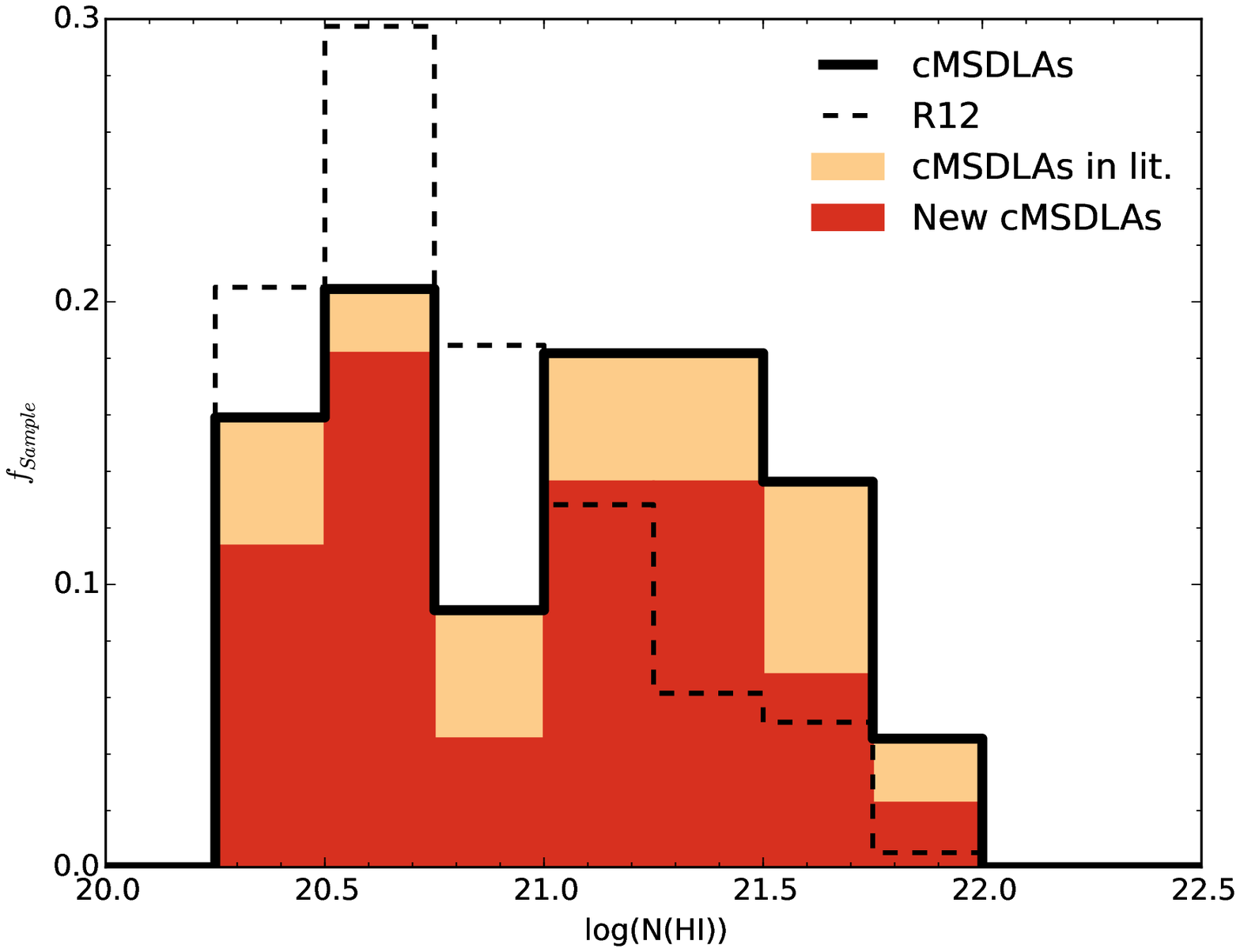}
\caption[N(\HI{}) distribution]{N(\HI{}) distribution of the cMSDLA sample (solid black line) compared to the R12 literature DLAs (black dashed line). The darker shade represents the fraction of cMSDLAs observed in this work, while the lighter shade shows the contribution from cMSDLAs already observed in the literature. Although the cMSDLAs span the entire range of N(\HI{}) values seen in the R12 DLAs; our sample is clearly biased towards systems with high \HI{} column densities, with nearly half having an \HI{} column density logN(\HI{}) $>$21.}
\label{fig:HIdist}
\end{center}
\end{figure}

\begin{figure}
\begin{center}
\includegraphics[width=\textwidth]{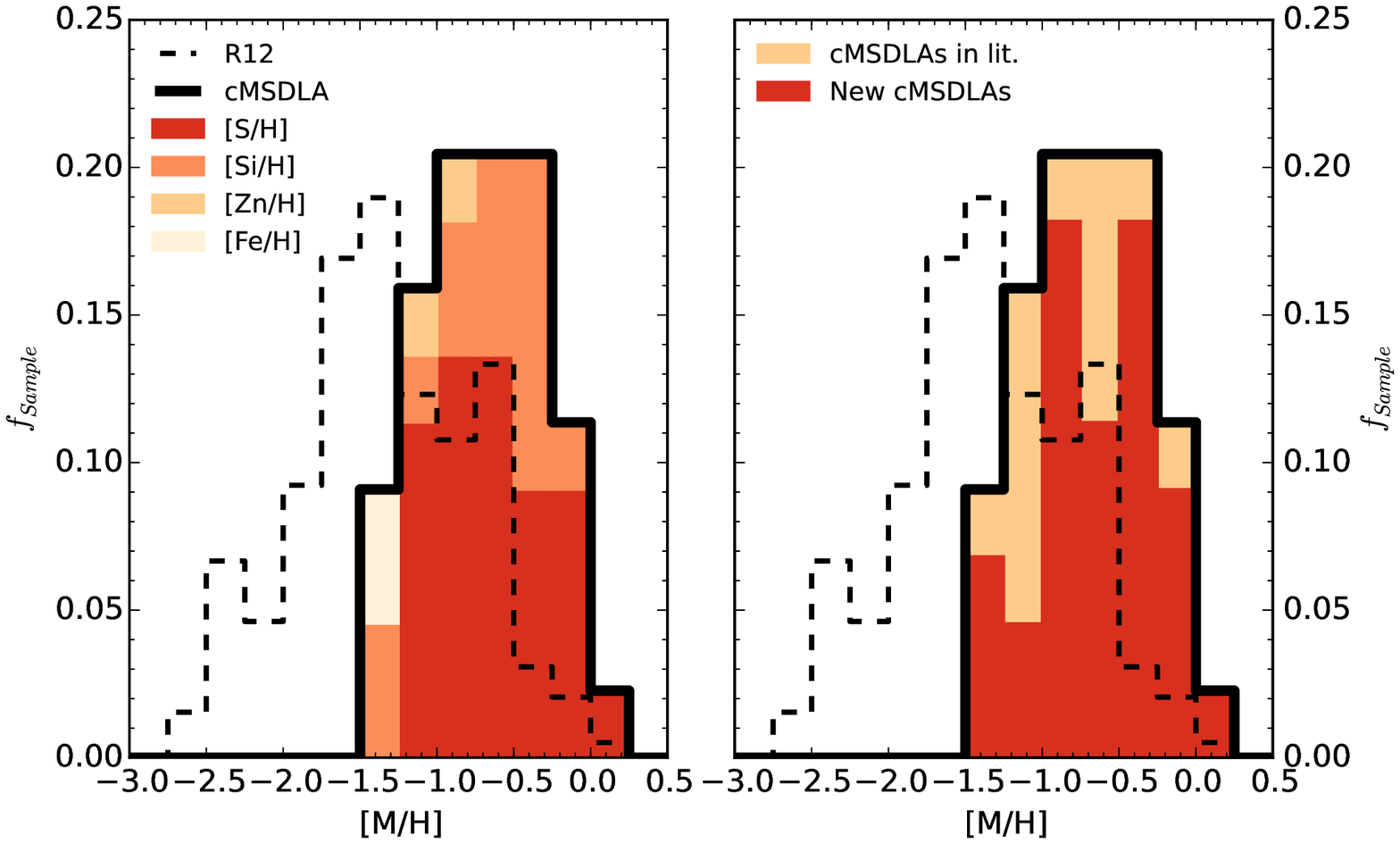}
\caption[Metallicity distribution]{Metallicity distribution of the cMSDLA sample compared to the R12 DLA sample (black dashed line). DLAs with limits on their metallicities are not shown. \emph{Left panel:} The different shades divide the distribution into fractions of which elements were used as a metallicity indicator in each bin. It is clear that cMSDLA sample probes a higher metallicity range than the DLAs from the R12 sample (median [M/H]$\sim-0.7$ dex compared to $\sim-1.5$ dex in the R12 sample). Sulphur is the dominant metallicity tracer adopted at higher metallicities (whereas silicon is the most used metallicity tracer in R12) as it is more likely reliably detected with higher column densities. \emph{Right panel:} The distribution is separated into the 13 cMSDLAs from the literature and the 31 new cMSDLAs, following the same shading scheme in Figure \ref{fig:HIdist}.}
\label{fig:metdist}
\end{center}
\end{figure}

\begin{figure}
\begin{center}
\includegraphics[width=\textwidth]{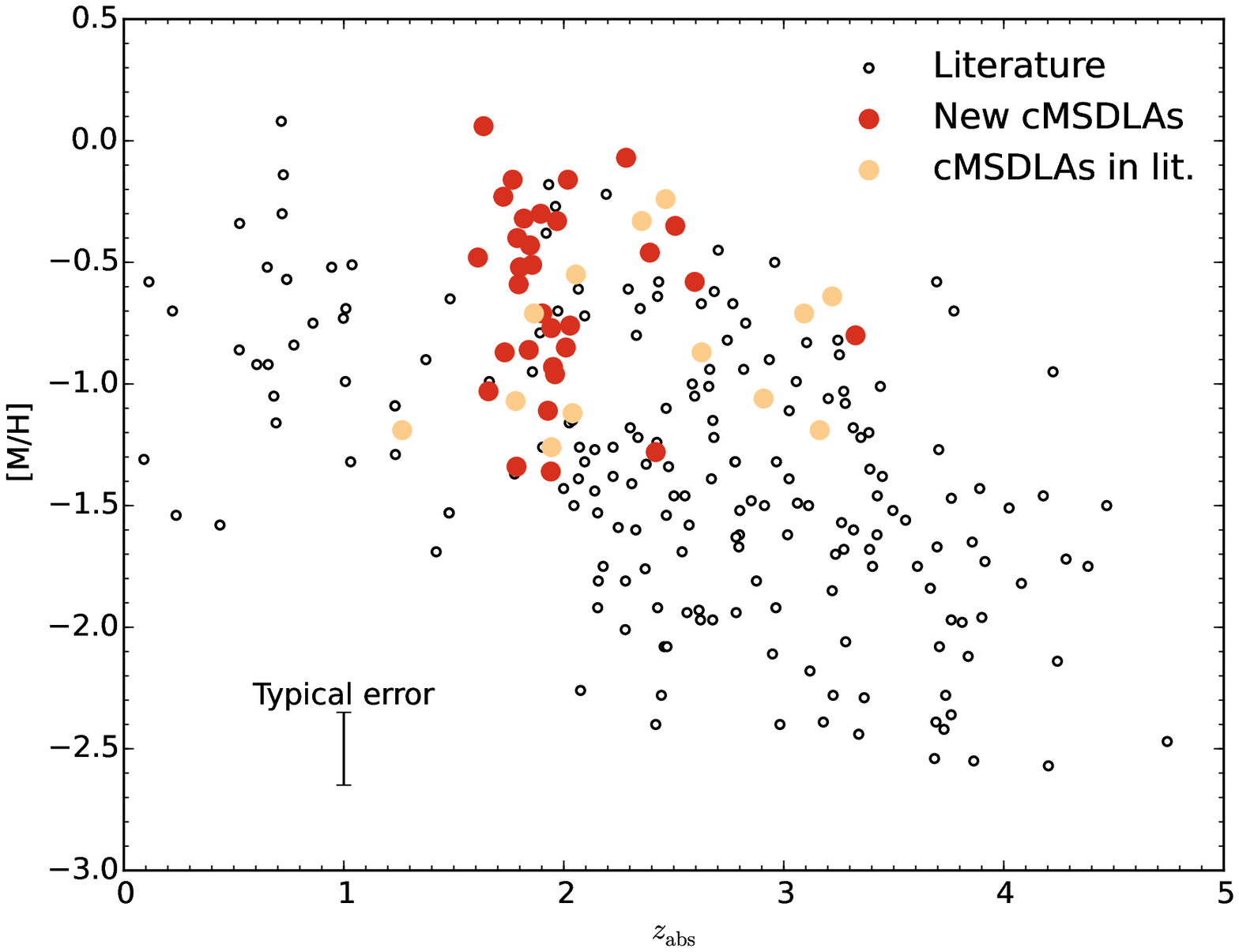}
\caption[Redshift distribution]{cMSDLA sample metallicity as a function of \zabs{}. The R12 sample DLAs are represented by the smaller unfilled circles, while the larger filled circles show the cMSDLA sample (darker shaded circles for the new cMSDLAs; lighter shaded circles for cMSDLAs taken from the literature). It is clear that the cMSDLA sample probes the highest metallicity DLAs within the redshift range of 1.5 $\lesssim$ \zabs{} $\lesssim$ $3.5$.}
\label{fig:zabmet}
\end{center}
\end{figure}

Figures \ref{fig:HIdist}, \ref{fig:metdist}, and \ref{fig:zabmet} show the distributions of neutral hydrogen column density, metallicity \citep[using the solar scale from][]{Asplund09}, and absorption redshift of our cMSDLA sample in comparison to a sample of DLAs from the literature compiled by R12. The R12 sample is a compilation of metal abundances of DLAs  published in the literature from studies that do not specifically target metal-poor or metal-strong systems. We have removed all the DLAs from the R12 sample that also appear in our cMSDLA sample. However, it is important to note that there are MSDLAs still present within the R12 sample after removing the duplicated DLAs from our sample ($\sim 6\%$; see discussion below).

Figure \ref{fig:HIdist} clearly shows that the distribution of neutral hydrogen column density for the cMSDLA sample spans the entire range of values as the DLAs in the R12 sample. However the cMSDLA sample shows a bias to higher N(\HI{}) systems in contrast to what is seen in the R12 sample  \citep[and other N(\HI{}) distributions seen for larger surveys, see][for example]{Noterdaeme12}. It is likely that due to selecting larger metal column densities,  high metal column density DLAs tend to probe more gas-rich systems and thus have higher \HI{} column densities relative to a typical DLA \citep{Kaplan10}. This excess of high N(\HI{}) DLAs is seen in Figure \ref{fig:HIdist}, where nearly half of the cMSDLAs in our sample have hydrogen column densities of log N(\HI{})$\sim21$, whereas the median \HI{} column density for the R12 literature sample is N(\HI{})$=20.7$. Furthermore, a KS test rules out that the two populations are drawn from the same parent sample at a 98.6\% confidence level. 

The metallicity distributions of the cMSDLA and R12 samples are shown in Figure \ref{fig:metdist}. Whereas the median metallicity of the R12 literature DLAs is [M/H]$=-1.51$, this value is the lowest metallicity of the cMSDLA sample distribution. We can safely say that the cMSDLA sample is indeed \emph{metal-rich} relative to the average DLA in the literature. Furthermore, the use of sulphur as a metallicity indicator is preferentially selected in the cMSDLA sample, as  the higher column densities allow for more frequent detections of the weaker S\sion{} lines. For comparison, the typical metallicity indicator in the R12 DLAs is Si\sion{}.

In terms of redshift (Figure \ref{fig:zabmet}), the cMSDLA sample does not span the entire range that the R12 literature DLA sample spans, but fall mostly within the redshift range of 1.5 to $3.5$.  This range in redshift is entirely due to selection effects; from choosing DLAs where the [C\sion{}] $\lambda$ 158 micron emission with ALMA, and observing both Ly-$\alpha$ and metal lines from ground--based observations without the atmospheric cutoffs of HIRES and SDSS \citep[HF06;][]{AdelmanMcCarthy08}. However, our cMSDLA sample probes the most metal-rich systems (compared to the R12 literature) for the redshifts in which they are observed.

\begin{figure}
\begin{center}
\includegraphics[width=\textwidth]{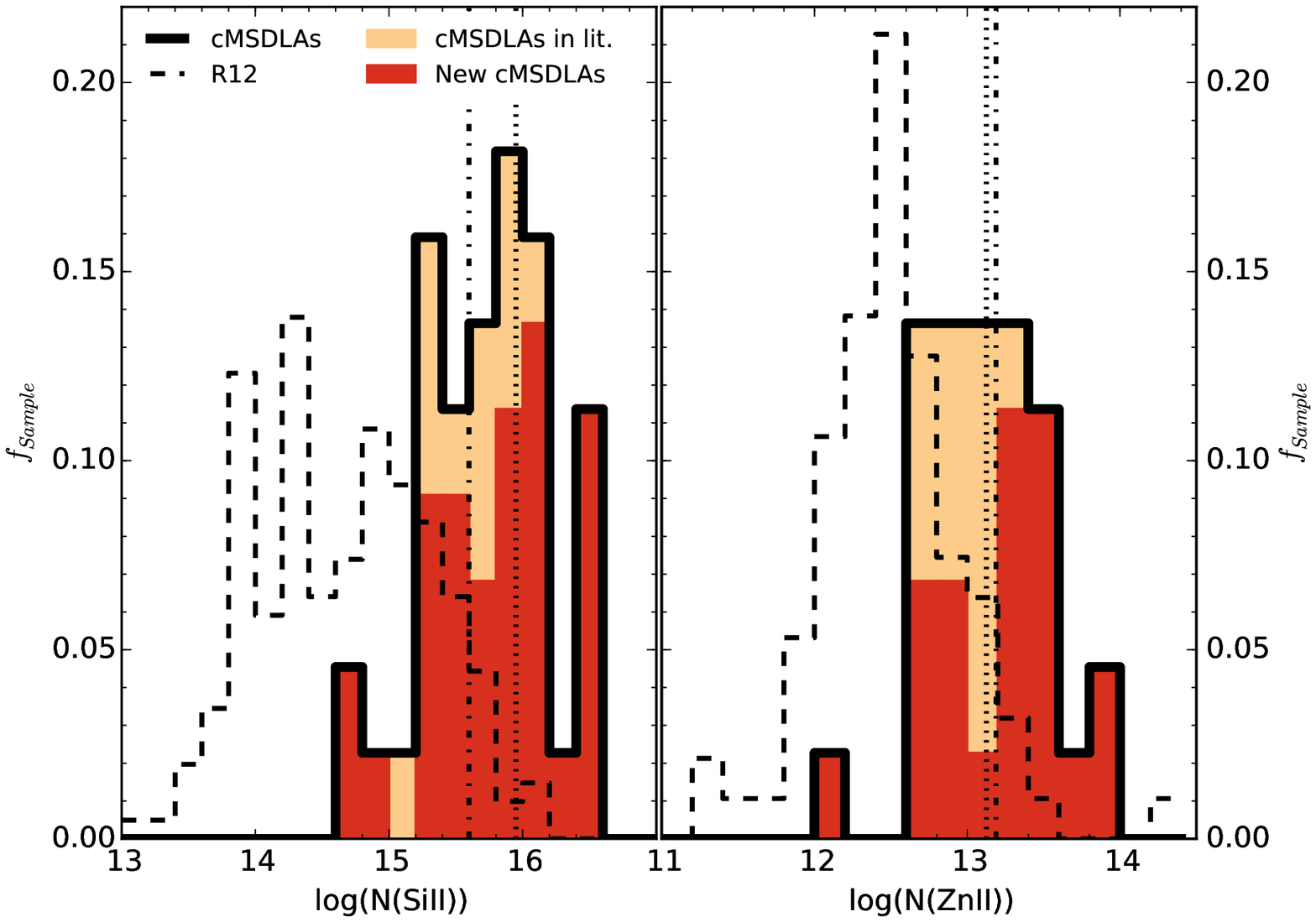}
\caption[N(Si\sion) and N(Zn\sion) distributions]{Column density distributions of Si\sion{} (\emph{left panel}) and Zn\sion{}  (\emph{right panel}) for both the cMSDLAs and DLAs within the R12 literature (following the same notation as in Figure \ref{fig:HIdist}). The MSDLA column density limits (N(Si\sion) $\geq$ 15.95; N(Zn\sion) $\geq$ 13.15) are shown as the vertical dotted lines in each panel. Only $\sim6$\% of DLAs observed in the R12 literature appear to be genuine MSDLAs; while $\sim45\%$ of the cMSDLA sample makes the HF06 column cut. Selecting the 90\ts{th} percentile of the R12 column density distributions (dotted-dashed lines) selects the majority of the cMSDLA sample. The 90\ts{th} percentile cut for logN(Zn) is nearly equivalent to the MSDLA column cut. For clarity, the two vertical lines in the right hand panel representing these column density cuts have been separated by 0.05 dex.}
\label{fig:NMSDLAdist}
\end{center}
\end{figure}

With the measured column densities from the previous section (Table \ref{tab:ColSumm}), the cMSDLA sample can be classified into bona fide MSDLAs or not. Using the column density cuts logN(Si\sion{}) $\geq$ 15.95 or logN(Zn\sion{}) $\geq$ 13.15 from HF06, 20 of our cMSDLA sample are truly MSDLAs, while the other 24 do not satisfy the requirements (presented in Table \ref{tab:DLAsum}). To highlight the number of DLAs that are true MSDLAs, Figure \ref{fig:NMSDLAdist} shows the distribution of column density distributions for Si\sion{} (left panel) and Zn\sion{} (right panel). For all DLAs in the R12 literature sample, only 6.6\% of DLAs make the MSDLA cut. The rarity of bona fide MSDLAs observed indicates that the column density cuts are targeting a very unique sample of DLAs.

It is interesting to note that there is little distinction between the metallicity distribution of bona fide MSDLAs and the DLAs from our sample that do not make the HF06 column density cuts, as inferred by their high N(\HI{}) and metal columns. In addition, other than in FJ0812+3208 (for which the HF06 MSDLA classification was defined), the only other $>$ $3\sigma$ detection of boron comes from a DLA that does not make the MSDLA cut defined in HF06 \citep{Berg13}. This suggests that the HF06 MSDLA classification scheme may be too strict for defining the optimal sample to study the nucleosynthesis of the most evolved galaxies at this epoch. The dashed--dotted line in Figure \ref{fig:NMSDLAdist} demonstrates that taking an arbitrary cut of the top 10\% of metal columns in the R12 literature sample (logN(Si) $>$ 15.60; logN(Zn) $>$ 13.15\footnote{Note that of the R12 literature DLAs where Zn\sion{} is detected, the 90\ts{th} percentile nearly corresponds  with the MSDLA cut in logN(Zn)$=13.15$. However, only 94 of the 260 R12 DLAs have a measured N(Zn). As zinc is a relatively rare element \citep[][]{Asplund09} and the Zn\sion $\lambda$ 2026 used for obtaining column densities is a naturally weak line, it is likely that these 94 systems are biased towards higher metal column densities where zinc can be detected. Therefore this 90\ts{th} percentile logN(Zn) cut is not representative of a general DLA distribution, and should be taken as an upper limit.}) can still provide an appropriate sample for the study of the most metal-rich systems in the early universe, despite not being true MSDLAs. We therefore do not further differentiate our cMSDLA sample between bona fide MSDLAs and DLAs that do not make the HF06 metal column density cuts, as we have demonstrated both types of DLAs are metal-rich (Figure \ref{fig:metdist}) and have probed the highest metal column density systems (Figure \ref{fig:NMSDLAdist}). Such a differentiation would have no overall effect to the science goals of (\emph{i}) searching for exotic elements in DLAs to provide nucleosynthetic constraints in the early universe, and (\emph{ii}) compare the chemistry of DLAs to the higher metallicity components of the Milky Way (i.e.~the thin and thick disk).

\subsection{Concluding Remarks}

In this paper we have added high resolution observations of an additional 31 candidate MSDLAs to a pre-existing sample of 13 systems. Our 44 system sample of metal-rich DLAs spans the entire range of \HI{} column densities seen in the R12 literature (although our sample is  biased to higher N(\HI{}) systems), and probes the higher end of the metallicity distribution of DLAs at redshifts $>$ 1.5. Furthermore, MSDLAs are very rare, accounting for only $\sim 6\%$ of all systems observed with high resolution spectrographs in the literature. However, we see little evidence for a fundamental difference between our DLA sample and DLAs that meet the subjective HF06 cut. We find that our entire sample remains useful for probing the top 10\% of metal column densities of DLAs (logN(Si) $>$ 15.60; logN(Zn) $>$ 13.15). With the typical redshift of our sample being at $z\sim2$, we are probing galactic chemical enrichment at times when the universe was only $\sim3$ Gyr old. As we will discuss in a forthcoming paper (Paper II), we can get a better understanding of chemical evolution history of DLAs in comparison to the different chemical regimes of the Milky Way system using this metal-rich DLA sample as they are nearly as enriched as the Milky Way already at $z\sim2$.

\section*{Acknowledgments}
\acknowledgments
We dedicate this paper in memory of A. M. Wolfe. We thank Marc Rafelski for providing us with the N(Si) and N(Zn) from his literature sample, and the anonymous referee for their useful comments on improving the manuscript. MN and JXP are partially supported by NSF grant AST-1109452. We wish to recognize and acknowledge the very significant cultural role and reverence that the summit of Mauna Kea has always had within the indigenous Hawaiian community.  We are most fortunate to have the opportunity to conduct observations from this mountain. The data presented herein were obtained at the W.M. Keck Observatory, which is operated as a scientific partnership among the California Institute of Technology, the University of California and the National Aeronautics and Space Administration. The Observatory was made possible by the generous financial support of the W.M. Keck Foundation.

\appendix
\section{Additional Material}

\setcounter{table}{0}
\renewcommand{\thetable}{\thesection.\arabic{table}}
 
\begin{table}
\begin{center}
\caption{Metal column densities for J0008$-$0958 (\zabs{}=$1.77$)}
\label{tab:J00080958}
\begin{tabular}{lcccccccc}
\hline
Ion& Line& $\lambda$& $f$& $v_{\rm min}$& $v_{\rm max}$& logN(X)& Included& logN$_{\rm adopt}$\\
 &  & \AA{}&  & \kms{}& \kms{}&  &  &  \\
\hline
Si\sion{}& 1260& 1260.422& $1.01\times 10^{0}$& $-245$& $150$& $> 14.51$& N&  \\
Si\sion{}& 1304& 1304.370& $9.40\times 10^{-2}$& $-245$& $150$& $> 15.38$& N&  \\
Si\sion{}& 1526& 1526.707& $1.27\times 10^{-1}$& $-245$& $150$& $> 15.22$& N&  \\
Si\sion{}& 1808& 1808.013& $2.19\times 10^{-3}$& $-245$& $150$& $16.04 \pm 0.05$& Y&  \\
Si\sion{}& --& --& --& --& --& --& --& $16.04 \pm 0.05$\\
\hline
S\sion{}& 1250& 1250.584& $5.43\times 10^{-3}$& $-245$& $150$& $15.84 \pm 0.05$& Y&  \\
\hline
\end{tabular}
\end{center}
\end{table}

\begin{table}
\begin{center}
\caption{Metal column densities for J0044+0018 (\zabs{}=$1.73$)}
\label{tab:J00440018}
\begin{tabular}{lcccccccc}
\hline
Ion& Line& $\lambda$& $f$& $v_{\rm min}$& $v_{\rm max}$& logN(X)& Included& logN$_{\rm adopt}$\\
 &  & \AA{}&  & \kms{}& \kms{}&  &  &  \\
\hline
Si\sion{}& 1304& 1304.370& $9.40\times 10^{-2}$& $-180$& $120$& $> 15.03$& N&  \\
Si\sion{}& 1526& 1526.707& $1.27\times 10^{-1}$& $-140$& $90$& $> 14.90$& N&  \\
Si\sion{}& 1808& 1808.013& $2.19\times 10^{-3}$& $-140$& $90$& $15.34 \pm 0.05$& Y&  \\
Si\sion{}& --& --& --& --& --& --& --& $15.34 \pm 0.05$\\
\hline
S\sion{}& 1250& 1250.584& $5.43\times 10^{-3}$& $-120$& $90$& $15.29 \pm 0.05$& Y&  \\
S\sion{}& 1253& 1253.811& $1.09\times 10^{-2}$& $-140$& $90$& $15.26 \pm 0.05$& Y&  \\
S\sion{}& --& --& --& --& --& --& --& $15.27 \pm 0.05$\\
\hline
Cr\sion{}& 2056& 2056.254& $1.05\times 10^{-1}$& $-140$& $70$& $< 13.04$& Y&  \\
Cr\sion{}& --& --& --& --& --& --& --& $< 13.04$\\
\hline
Fe\sion{}& 1608& 1608.451& $5.80\times 10^{-2}$& $-140$& $90$& $> 14.77$& Y&  \\
Fe\sion{}& --& --& --& --& --& --& --& $> 14.77$\\
\hline
Ni\sion{}& 1317& 1317.217& $5.71\times 10^{-2}$& $-140$& $90$& $13.94 \pm 0.05$& Y&  \\
Ni\sion{}& 1741& 1741.553& $4.27\times 10^{-2}$& $-140$& $90$& $13.83 \pm 0.06$& Y&  \\
Ni\sion{}& --& --& --& --& --& --& --& $13.89 \pm 0.05$\\
\hline
Zn\sion{}& 2026& 2026.136& $4.89\times 10^{-1}$& $-140$& $90$& $< 12.61$& Y&  \\
Zn\sion{}& --& --& --& --& --& --& --& $< 12.61$\\
\hline
\end{tabular}
\end{center}
\end{table}

\begin{table}
\begin{center}
\caption{Metal column densities for J0058+0115 (\zabs{}=$2.01$)}
\label{tab:J00580115}
\begin{tabular}{lcccccccc}
\hline
Ion& Line& $\lambda$& $f$& $v_{\rm min}$& $v_{\rm max}$& logN(X)& Included& logN$_{\rm adopt}$\\
 &  & \AA{}&  & \kms{}& \kms{}&  &  &  \\
\hline
Si\sion{}& 1260& 1260.422& $1.01\times 10^{0}$& $-90$& $400$& $> 14.47$& N&  \\
Si\sion{}& 1304& 1304.370& $9.40\times 10^{-2}$& $-90$& $400$& $> 15.29$& Y&  \\
Si\sion{}& 1526& 1526.707& $1.27\times 10^{-1}$& $-90$& $400$& $> 15.16$& Y&  \\
Si\sion{}& 1808& 1808.013& $2.19\times 10^{-3}$& $-90$& $80$& $> 15.55$& Y&  \\
Si\sion{}& --& --& --& --& --& --& --& $> 15.55$\\
\hline
S\sion{}& 1253& 1253.811& $1.09\times 10^{-2}$& $-90$& $400$& $15.40 \pm 0.05$& Y&  \\
S\sion{}& --& --& --& --& --& --& --& $15.40 \pm 0.05$\\
\hline
Cr\sion{}& 2056& 2056.254& $1.05\times 10^{-1}$& $-60$& $300$& $13.54 \pm 0.05$& Y&  \\
Cr\sion{}& --& --& --& --& --& --& --& $13.54 \pm 0.05$\\
\hline
Fe\sion{}& 1608& 1608.451& $5.80\times 10^{-2}$& $-90$& $400$& $> 15.19$& N&  \\
Fe\sion{}& 1611& 1611.200& $1.36\times 10^{-3}$& $-90$& $400$& $15.18 \pm 0.05$& Y&  \\
Fe\sion{}& --& --& --& --& --& --& --& $15.18 \pm 0.05$\\
\hline
Ni\sion{}& 1709& 1709.604& $3.24\times 10^{-2}$& $-90$& $400$& $14.10 \pm 0.05$& Y&  \\
Ni\sion{}& 1741& 1741.553& $4.27\times 10^{-2}$& $-90$& $400$& $14.19 \pm 0.05$& Y&  \\
Ni\sion{}& 1751& 1751.916& $2.77\times 10^{-2}$& $-90$& $400$& $14.07 \pm 0.06$& Y&  \\
Ni\sion{}& --& --& --& --& --& --& --& $14.16 \pm 0.05$\\
\hline
Zn\sion{}& 2026& 2026.136& $4.89\times 10^{-1}$& $-60$& $400$& $12.95 \pm 0.05$& Y&  \\
Zn\sion{}& --& --& --& --& --& --& --& $12.95 \pm 0.05$\\
\hline
\end{tabular}
\end{center}
\end{table}

\begin{table}
\begin{center}
\caption{Metal column densities for J0211+1241 (\zabs{}=$2.60$)}
\label{tab:J02111241}
\begin{tabular}{lcccccccc}
\hline
Ion& Line& $\lambda$& $f$& $v_{\rm min}$& $v_{\rm max}$& logN(X)& Included& logN$_{\rm adopt}$\\
 &  & \AA{}&  & \kms{}& \kms{}&  &  &  \\
\hline
Si\sion{}& 1526& 1526.707& $1.27\times 10^{-1}$& $-470$& $90$& $> 14.89$& Y&  \\
Si\sion{}& 1808& 1808.013& $2.19\times 10^{-3}$& $-470$& $90$& $15.53 \pm 0.08$& Y&  \\
Si\sion{}& --& --& --& --& --& --& --& $15.53 \pm 0.08$\\
\hline
Fe\sion{}& 1143& 1143.226& $1.77\times 10^{-2}$& $-80$& $40$& $15.06 \pm 0.05$& Y&  \\
Fe\sion{}& 1608& 1608.451& $5.80\times 10^{-2}$& $-470$& $90$& $> 14.90$& Y&  \\
Fe\sion{}& 1611& 1611.200& $1.36\times 10^{-3}$& $-470$& $90$& $< 16.05$& Y&  \\
Fe\sion{}& --& --& --& --& --& --& --& $15.06 \pm 0.05$\\
\hline
Ni\sion{}& 1709& 1709.604& $3.24\times 10^{-2}$& $-470$& $90$& $14.07 \pm 0.06$& Y&  \\
Ni\sion{}& 1741& 1741.553& $4.27\times 10^{-2}$& $-470$& $90$& $14.06 \pm 0.10$& Y&  \\
Ni\sion{}& 1751& 1751.916& $2.77\times 10^{-2}$& $-470$& $90$& $< 14.25$& N&  \\
Ni\sion{}& --& --& --& --& --& --& --& $14.07 \pm 0.05$\\
\hline
\end{tabular}
\end{center}
\end{table}

\begin{table}
\begin{center}
\caption{Metal column densities for J0233+0103 (\zabs{}=$1.78$)}
\label{tab:J02330103}
\begin{tabular}{lcccccccc}
\hline
Ion& Line& $\lambda$& $f$& $v_{\rm min}$& $v_{\rm max}$& logN(X)& Included& logN$_{\rm adopt}$\\
 &  & \AA{}&  & \kms{}& \kms{}&  &  &  \\
\hline
Si\sion{}& 1304& 1304.370& $9.40\times 10^{-2}$& $-100$& $60$& $> 14.78$& Y&  \\
Si\sion{}& 1526& 1526.707& $1.27\times 10^{-1}$& $-100$& $60$& $> 14.69$& Y&  \\
Si\sion{}& 1808& 1808.013& $2.19\times 10^{-3}$& $-100$& $60$& $< 14.76$& Y&  \\
Si\sion{}& --& --& --& --& --& --& --& $14.77 \pm 0.05$\\
\hline
Fe\sion{}& 1608& 1608.451& $5.80\times 10^{-2}$& $-100$& $60$& $14.62 \pm 0.05$& Y&  \\
Fe\sion{}& --& --& --& --& --& --& --& $14.62 \pm 0.05$\\
\hline
Ni\sion{}& 1741& 1741.553& $4.27\times 10^{-2}$& $-100$& $60$& $13.61 \pm 0.11$& Y&  \\
Ni\sion{}& --& --& --& --& --& --& --& $13.61 \pm 0.11$\\
\hline
\end{tabular}
\end{center}
\end{table}

\begin{table}
\begin{center}
\caption{Metal column densities for J0815+1037 (\zabs{}=$1.85$)}
\label{tab:J08151037}
\begin{tabular}{lcccccccc}
\hline
Ion& Line& $\lambda$& $f$& $v_{\rm min}$& $v_{\rm max}$& logN(X)& Included& logN$_{\rm adopt}$\\
 &  & \AA{}&  & \kms{}& \kms{}&  &  &  \\
\hline
Si\sion{}& 1304& 1304.370& $9.40\times 10^{-2}$& $-200$& $100$& $> 14.95$& Y&  \\
Si\sion{}& 1526& 1526.707& $1.27\times 10^{-1}$& $-200$& $100$& $> 14.73$& Y&  \\
Si\sion{}& 1808& 1808.013& $2.19\times 10^{-3}$& $-200$& $100$& $15.38 \pm 0.45$& Y&  \\
Si\sion{}& --& --& --& --& --& --& --& $15.38 \pm 0.45$\\
\hline
Fe\sion{}& 1608& 1608.451& $5.80\times 10^{-2}$& $-200$& $100$& $> 14.87$& Y&  \\
Fe\sion{}& --& --& --& --& --& --& --& $> 14.87$\\
\hline
Ni\sion{}& 1317& 1317.217& $5.71\times 10^{-2}$& $-50$& $50$& $13.74 \pm 0.12$& Y&  \\
Ni\sion{}& --& --& --& --& --& --& --& $13.74 \pm 0.12$\\
\hline
\end{tabular}
\end{center}
\end{table}

\begin{table}
\begin{center}
\caption{Metal column densities for J0927+1543 (\zabs{}=$1.73$)}
\label{tab:J09271543}
\begin{tabular}{lcccccccc}
\hline
Ion& Line& $\lambda$& $f$& $v_{\rm min}$& $v_{\rm max}$& logN(X)& Included& logN$_{\rm adopt}$\\
 &  & \AA{}&  & \kms{}& \kms{}&  &  &  \\
\hline
Si\sion{}& 1304& 1304.370& $9.40\times 10^{-2}$& $-230$& $190$& $> 15.27$& N&  \\
Si\sion{}& 1526& 1526.707& $1.27\times 10^{-1}$& $-230$& $190$& $> 15.14$& N&  \\
Si\sion{}& 1808& 1808.013& $2.19\times 10^{-3}$& $-230$& $190$& $15.99 \pm 0.05$& Y&  \\
Si\sion{}& --& --& --& --& --& --& --& $15.99 \pm 0.05$\\
\hline
Cr\sion{}& 2056& 2056.254& $1.05\times 10^{-1}$& $-230$& $190$& $13.86 \pm 0.05$& Y&  \\
Cr\sion{}& 2066& 2066.161& $5.15\times 10^{-2}$& $-230$& $190$& $13.76 \pm 0.09$& Y&  \\
Cr\sion{}& --& --& --& --& --& --& --& $13.83 \pm 0.05$\\
\hline
Fe\sion{}& 1608& 1608.451& $5.80\times 10^{-2}$& $-230$& $190$& $> 15.17$& Y&  \\
Fe\sion{}& 1611& 1611.200& $1.36\times 10^{-3}$& $-230$& $190$& $< 15.13$& Y&  \\
Fe\sion{}& --& --& --& --& --& --& --& $15.14 \pm 0.24$\\
\hline
Ni\sion{}& 1709& 1709.604& $3.24\times 10^{-2}$& $-230$& $190$& $14.04 \pm 0.09$& Y&  \\
Ni\sion{}& 1741& 1741.553& $4.27\times 10^{-2}$& $-230$& $190$& $14.19 \pm 0.05$& Y&  \\
Ni\sion{}& 1751& 1751.916& $2.77\times 10^{-2}$& $-230$& $190$& $14.26 \pm 0.06$& Y&  \\
Ni\sion{}& --& --& --& --& --& --& --& $14.17 \pm 0.05$\\
\hline
Zn\sion{}& 2026& 2026.136& $4.89\times 10^{-1}$& $-230$& $190$& $< 13.69$& Y&  \\
Zn\sion{}& 2062& 2062.664& $2.56\times 10^{-1}$& $-230$& $190$& $13.38 \pm 0.05$& Y&  \\
Zn\sion{}& --& --& --& --& --& --& --& $13.38 \pm 0.05$\\
\hline
\end{tabular}
\end{center}
\end{table}

\begin{table}
\begin{center}
\caption{Metal column densities for J0927+5823 (\zabs{}=$1.64$)}
\label{tab:J09275823}
\begin{tabular}{lcccccccc}
\hline
Ion& Line& $\lambda$& $f$& $v_{\rm min}$& $v_{\rm max}$& logN(X)& Included& logN$_{\rm adopt}$\\
 &  & \AA{}&  & \kms{}& \kms{}&  &  &  \\
\hline
Si\sion{}& 1526& 1526.707& $1.27\times 10^{-1}$& $-255$& $255$& $> 15.24$& N&  \\
Si\sion{}& 1808& 1808.013& $2.19\times 10^{-3}$& $-255$& $255$& $15.72 \pm 0.05$& Y&  \\
Si\sion{}& --& --& --& --& --& --& --& $15.72 \pm 0.05$\\
\hline
S\sion{}& 1250& 1250.584& $5.43\times 10^{-3}$& $-130$& $180$& $15.61 \pm 0.05$& Y&  \\
S\sion{}& --& --& --& --& --& --& --& $15.61 \pm 0.05$\\
\hline
Cr\sion{}& 2056& 2056.254& $1.05\times 10^{-1}$& $-130$& $255$& $13.55 \pm 0.05$& N&  \\
Cr\sion{}& --& --& --& --& --& --& --& \nodata{}\\
\hline
Fe\sion{}& 1608& 1608.451& $5.80\times 10^{-2}$& $-255$& $255$& $> 15.27$& Y&  \\
Fe\sion{}& 1611& 1611.200& $1.36\times 10^{-3}$& $-255$& $255$& $< 14.95$& Y&  \\
Fe\sion{}& --& --& --& --& --& --& --& $> 15.27$\\
\hline
Ni\sion{}& 1317& 1317.217& $5.71\times 10^{-2}$& $-255$& $255$& $14.45 \pm 0.05$& Y&  \\
Ni\sion{}& 1454& 1454.842& $3.23\times 10^{-2}$& $-255$& $255$& $14.45 \pm 0.05$& Y&  \\
Ni\sion{}& 1709& 1709.604& $3.24\times 10^{-2}$& $-255$& $255$& $14.46 \pm 0.05$& Y&  \\
Ni\sion{}& 1741& 1741.553& $4.27\times 10^{-2}$& $-255$& $255$& $14.42 \pm 0.05$& Y&  \\
Ni\sion{}& --& --& --& --& --& --& --& $14.44 \pm 0.05$\\
\hline
Zn\sion{}& 2026& 2026.136& $4.89\times 10^{-1}$& $-130$& $180$& $13.27 \pm 0.05$& Y&  \\
Zn\sion{}& 2062& 2062.664& $2.56\times 10^{-1}$& $-130$& $130$& $13.38 \pm 0.05$& Y&  \\
Zn\sion{}& --& --& --& --& --& --& --& $13.29 \pm 0.05$\\
\hline
\end{tabular}
\end{center}
\end{table}

\begin{table}
\begin{center}
\caption{Metal column densities for J0958+0145 (\zabs{}=$1.93$)}
\label{tab:J09580145}
\begin{tabular}{lcccccccc}
\hline
Ion& Line& $\lambda$& $f$& $v_{\rm min}$& $v_{\rm max}$& logN(X)& Included& logN$_{\rm adopt}$\\
 &  & \AA{}&  & \kms{}& \kms{}&  &  &  \\
\hline
Si\sion{}& 1260& 1260.422& $1.01\times 10^{0}$& $-120$& $100$& $> 14.07$& N&  \\
Si\sion{}& 1304& 1304.370& $9.40\times 10^{-2}$& $-120$& $100$& $> 14.65$& N&  \\
Si\sion{}& 1526& 1526.707& $1.27\times 10^{-1}$& $-120$& $100$& $> 14.54$& N&  \\
Si\sion{}& 1808& 1808.013& $2.19\times 10^{-3}$& $-25$& $60$& $14.84 \pm 0.06$& Y&  \\
Si\sion{}& --& --& --& --& --& --& --& $14.84 \pm 0.06$\\
\hline
S\sion{}& 1250& 1250.584& $5.43\times 10^{-3}$& $-25$& $60$& $14.49 \pm 0.05$& Y&  \\
S\sion{}& 1253& 1253.811& $1.09\times 10^{-2}$& $-25$& $60$& $14.37 \pm 0.05$& Y&  \\
S\sion{}& 1259& 1259.519& $1.66\times 10^{-2}$& $-25$& $60$& $14.47 \pm 0.05$& Y&  \\
S\sion{}& --& --& --& --& --& --& --& $14.44 \pm 0.05$\\
\hline
Fe\sion{}& 1121& 1121.975& $2.02\times 10^{-2}$& $-25$& $60$& $14.08 \pm 0.14$& Y&  \\
Fe\sion{}& 1133& 1133.665& $5.50\times 10^{-3}$& $-25$& $60$& $14.45 \pm 0.12$& Y&  \\
Fe\sion{}& 1143& 1143.226& $1.77\times 10^{-2}$& $-25$& $60$& $14.25 \pm 0.06$& Y&  \\
Fe\sion{}& 1608& 1608.451& $5.80\times 10^{-2}$& $-25$& $60$& $> 14.25$& Y&  \\
Fe\sion{}& 1611& 1611.200& $1.36\times 10^{-3}$& $-25$& $60$& $< 14.60$& Y&  \\
Fe\sion{}& --& --& --& --& --& --& --& $14.23 \pm 0.05$\\
\hline
Ni\sion{}& 1317& 1317.217& $5.71\times 10^{-2}$& $-25$& $60$& $13.36 \pm 0.08$& Y&  \\
Ni\sion{}& 1370& 1370.131& $5.88\times 10^{-2}$& $-25$& $50$& $> 13.05$& Y&  \\
Ni\sion{}& 1709& 1709.604& $3.24\times 10^{-2}$& $-25$& $60$& $13.40 \pm 0.11$& Y&  \\
Ni\sion{}& --& --& --& --& --& --& --& $13.37 \pm 0.07$\\
\hline
Zn\sion{}& 2026& 2026.136& $4.89\times 10^{-1}$& $-25$& $60$& $< 12.00$& Y&  \\
Zn\sion{}& --& --& --& --& --& --& --& $< 12.00$\\
\hline
\end{tabular}
\end{center}
\end{table}

\begin{table}
\begin{center}
\caption{Metal column densities for J1013+5615 (\zabs{}=$2.28$)}
\label{tab:J10135615}
\begin{tabular}{lcccccccc}
\hline
Ion& Line& $\lambda$& $f$& $v_{\rm min}$& $v_{\rm max}$& logN(X)& Included& logN$_{\rm adopt}$\\
 &  & \AA{}&  & \kms{}& \kms{}&  &  &  \\
\hline
Si\sion{}& 1526& 1526.707& $1.27\times 10^{-1}$& $-140$& $185$& $> 15.21$& N&  \\
Si\sion{}& 1808& 1808.013& $2.19\times 10^{-3}$& $-140$& $175$& $16.14 \pm 0.05$& Y&  \\
Si\sion{}& --& --& --& --& --& --& --& $16.14 \pm 0.05$\\
\hline
Cr\sion{}& 2056& 2056.254& $1.05\times 10^{-1}$& $-140$& $175$& $13.80 \pm 0.05$& Y&  \\
Cr\sion{}& 2066& 2066.161& $5.15\times 10^{-2}$& $-140$& $175$& $13.66 \pm 0.08$& Y&  \\
Cr\sion{}& --& --& --& --& --& --& --& $13.79 \pm 0.05$\\
\hline
Fe\sion{}& 1608& 1608.451& $5.80\times 10^{-2}$& $-140$& $175$& $> 15.44$& Y&  \\
Fe\sion{}& --& --& --& --& --& --& --& $> 15.45$\\
\hline
Zn\sion{}& 2026& 2026.136& $4.89\times 10^{-1}$& $-140$& $175$& $13.56 \pm 0.05$& Y&  \\
Zn\sion{}& --& --& --& --& --& --& --& $13.56 \pm 0.05$\\
\hline
\end{tabular}
\end{center}
\end{table}

\begin{table}
\begin{center}
\caption{Metal column densities for J1024+0600 (\zabs{}=$1.90$)}
\label{tab:J10240600}
\begin{tabular}{lcccccccc}
\hline
Ion& Line& $\lambda$& $f$& $v_{\rm min}$& $v_{\rm max}$& logN(X)& Included& logN$_{\rm adopt}$\\
 &  & \AA{}&  & \kms{}& \kms{}&  &  &  \\
\hline
Si\sion{}& 1304& 1304.370& $9.40\times 10^{-2}$& $-250$& $220$& $> 15.33$& N&  \\
Si\sion{}& 1526& 1526.707& $1.27\times 10^{-1}$& $-250$& $220$& $> 15.19$& N&  \\
Si\sion{}& 1808& 1808.013& $2.19\times 10^{-3}$& $-90$& $160$& $15.81 \pm 0.05$& Y&  \\
Si\sion{}& --& --& --& --& --& --& --& $15.81 \pm 0.05$\\
\hline
S\sion{}& 1250& 1250.584& $5.43\times 10^{-3}$& $-60$& $160$& $15.42 \pm 0.05$& Y&  \\
S\sion{}& 1253& 1253.811& $1.09\times 10^{-2}$& $-60$& $160$& $15.47 \pm 0.05$& Y&  \\
S\sion{}& --& --& --& --& --& --& --& $15.45 \pm 0.05$\\
\hline
Fe\sion{}& 1608& 1608.451& $5.80\times 10^{-2}$& $-250$& $160$& $> 15.21$& N&  \\
Fe\sion{}& 1611& 1611.200& $1.36\times 10^{-3}$& $-60$& $50$& $15.27 \pm 0.08$& Y&  \\
Fe\sion{}& --& --& --& --& --& --& --& $15.27 \pm 0.08$\\
\hline
Ni\sion{}& 1317& 1317.217& $5.71\times 10^{-2}$& $-60$& $140$& $13.93 \pm 0.05$& Y&  \\
Ni\sion{}& 1370& 1370.131& $5.88\times 10^{-2}$& $-60$& $160$& $14.06 \pm 0.05$& Y&  \\
Ni\sion{}& 1709& 1709.604& $3.24\times 10^{-2}$& $-60$& $160$& $14.26 \pm 0.05$& Y&  \\
Ni\sion{}& 1741& 1741.553& $4.27\times 10^{-2}$& $-60$& $160$& $14.20 \pm 0.05$& Y&  \\
Ni\sion{}& --& --& --& --& --& --& --& $14.02 \pm 0.05$\\
\hline
\end{tabular}
\end{center}
\end{table}

\begin{table}
\begin{center}
\caption{Metal column densities for J1042+0628 (\zabs{}=$1.94$)}
\label{tab:J10420628}
\begin{tabular}{lcccccccc}
\hline
Ion& Line& $\lambda$& $f$& $v_{\rm min}$& $v_{\rm max}$& logN(X)& Included& logN$_{\rm adopt}$\\
 &  & \AA{}&  & \kms{}& \kms{}&  &  &  \\
\hline
Si\sion{}& 1260& 1260.422& $1.01\times 10^{0}$& $-140$& $360$& $> 14.45$& N&  \\
Si\sion{}& 1304& 1304.370& $9.40\times 10^{-2}$& $-140$& $360$& $> 15.08$& N&  \\
Si\sion{}& 1526& 1526.707& $1.27\times 10^{-1}$& $-140$& $360$& $> 15.01$& N&  \\
Si\sion{}& 1808& 1808.013& $2.19\times 10^{-3}$& $-100$& $150$& $15.40 \pm 0.08$& Y&  \\
Si\sion{}& --& --& --& --& --& --& --& $15.40 \pm 0.08$\\
\hline
S\sion{}& 1250& 1250.584& $5.43\times 10^{-3}$& $-100$& $150$& $14.89 \pm 0.06$& Y&  \\
S\sion{}& 1253& 1253.811& $1.09\times 10^{-2}$& $-100$& $150$& $15.13 \pm 0.05$& Y&  \\
S\sion{}& --& --& --& --& --& --& --& $15.08 \pm 0.05$\\
\hline
Fe\sion{}& 1608& 1608.451& $5.80\times 10^{-2}$& $-140$& $360$& $> 14.86$& Y&  \\
Fe\sion{}& 1611& 1611.200& $1.36\times 10^{-3}$& $-100$& $150$& $< 15.13$& Y&  \\
Fe\sion{}& --& --& --& --& --& --& --& $15.00 \pm 0.15$\\
\hline
Ni\sion{}& 1317& 1317.217& $5.71\times 10^{-2}$& $-40$& $150$& $< 13.82$& Y&  \\
Ni\sion{}& 1370& 1370.131& $5.88\times 10^{-2}$& $-100$& $150$& $< 13.84$& Y&  \\
Ni\sion{}& 1709& 1709.604& $3.24\times 10^{-2}$& $-100$& $150$& $< 13.78$& Y&  \\
Ni\sion{}& --& --& --& --& --& --& --& $< 13.78$\\
\hline
\end{tabular}
\end{center}
\end{table}

\begin{table}
\begin{center}
\caption{Metal column densities for J1049$-$0110 (\zabs{}=$1.66$)}
\label{tab:J10490110}
\begin{tabular}{lcccccccc}
\hline
Ion& Line& $\lambda$& $f$& $v_{\rm min}$& $v_{\rm max}$& logN(X)& Included& logN$_{\rm adopt}$\\
 &  & \AA{}&  & \kms{}& \kms{}&  &  &  \\
\hline
Si\sion{}& 1304& 1304.370& $9.40\times 10^{-2}$& $-190$& $400$& $> 15.38$& N&  \\
Si\sion{}& 1526& 1526.707& $1.27\times 10^{-1}$& $-190$& $400$& $> 15.24$& Y&  \\
Si\sion{}& 1808& 1808.013& $2.19\times 10^{-3}$& $-190$& $400$& $15.80 \pm 0.05$& Y&  \\
Si\sion{}& --& --& --& --& --& --& --& $15.80 \pm 0.05$\\
\hline
S\sion{}& 1250& 1250.584& $5.43\times 10^{-3}$& $-10$& $400$& $< 15.47$& N&  \\
S\sion{}& 1253& 1253.811& $1.09\times 10^{-2}$& $-190$& $400$& $15.47 \pm 0.05$& Y&  \\
S\sion{}& --& --& --& --& --& --& --& $15.47 \pm 0.05$\\
\hline
Cr\sion{}& 2056& 2056.254& $1.05\times 10^{-1}$& $-70$& $400$& $13.48 \pm 0.05$& Y&  \\
Cr\sion{}& 2066& 2066.161& $5.15\times 10^{-2}$& $-100$& $200$& $13.51 \pm 0.07$& Y&  \\
Cr\sion{}& --& --& --& --& --& --& --& $13.49 \pm 0.05$\\
\hline
Fe\sion{}& 1608& 1608.451& $5.80\times 10^{-2}$& $-190$& $400$& $> 15.15$& N&  \\
Fe\sion{}& 1611& 1611.200& $1.36\times 10^{-3}$& $-120$& $280$& $< 14.97$& N&  \\
Fe\sion{}& 2260& 2260.780& $2.44\times 10^{-3}$& $-90$& $200$& $15.17 \pm 0.05$& Y&  \\
Fe\sion{}& --& --& --& --& --& --& --& $15.17 \pm 0.05$\\
\hline
Ni\sion{}& 1317& 1317.217& $5.71\times 10^{-2}$& $-70$& $300$& $14.29 \pm 0.05$& Y&  \\
Ni\sion{}& 1454& 1454.842& $3.23\times 10^{-2}$& $-190$& $400$& $13.89 \pm 0.08$& Y&  \\
Ni\sion{}& 1709& 1709.604& $3.24\times 10^{-2}$& $-190$& $400$& $14.38 \pm 0.05$& Y&  \\
Ni\sion{}& 1741& 1741.553& $4.27\times 10^{-2}$& $-20$& $400$& $14.30 \pm 0.05$& Y&  \\
Ni\sion{}& 1751& 1751.916& $2.77\times 10^{-2}$& $-190$& $400$& $14.16 \pm 0.05$& Y&  \\
Ni\sion{}& --& --& --& --& --& --& --& $14.25 \pm 0.05$\\
\hline
Zn\sion{}& 2026& 2026.136& $4.89\times 10^{-1}$& $-190$& $400$& $13.14 \pm 0.05$& Y&  \\
Zn\sion{}& --& --& --& --& --& --& --& $13.14 \pm 0.05$\\
\hline
\end{tabular}
\end{center}
\end{table}

\begin{table}
\begin{center}
\caption{Metal column densities for J1056+1208 (\zabs{}=$1.61$)}
\label{tab:J10561208}
\begin{tabular}{lcccccccc}
\hline
Ion& Line& $\lambda$& $f$& $v_{\rm min}$& $v_{\rm max}$& logN(X)& Included& logN$_{\rm adopt}$\\
 &  & \AA{}&  & \kms{}& \kms{}&  &  &  \\
\hline
Si\sion{}& 1304& 1304.370& $9.40\times 10^{-2}$& $-190$& $210$& $> 15.28$& Y&  \\
Si\sion{}& 1526& 1526.707& $1.27\times 10^{-1}$& $-190$& $210$& $> 15.09$& Y&  \\
Si\sion{}& 1808& 1808.013& $2.19\times 10^{-3}$& $-190$& $210$& $16.48 \pm 0.09$& Y&  \\
Si\sion{}& --& --& --& --& --& --& --& $16.48 \pm 0.09$\\
\hline
S\sion{}& 1250& 1250.584& $5.43\times 10^{-3}$& $-190$& $210$& $> 16.15$& Y&  \\
S\sion{}& --& --& --& --& --& --& --& $> 16.15$\\
\hline
\end{tabular}
\end{center}
\end{table}

\begin{table}
\begin{center}
\caption{Metal column densities for J1106+1044 (\zabs{}=$1.82$)}
\label{tab:J11061044}
\begin{tabular}{lcccccccc}
\hline
Ion& Line& $\lambda$& $f$& $v_{\rm min}$& $v_{\rm max}$& logN(X)& Included& logN$_{\rm adopt}$\\
 &  & \AA{}&  & \kms{}& \kms{}&  &  &  \\
\hline
Si\sion{}& 1304& 1304.370& $9.40\times 10^{-2}$& $-220$& $100$& $> 15.22$& Y&  \\
Si\sion{}& 1526& 1526.707& $1.27\times 10^{-1}$& $-220$& $100$& $> 15.11$& Y&  \\
Si\sion{}& --& --& --& --& --& --& --& $> 15.22$\\
\hline
S\sion{}& 1250& 1250.584& $5.43\times 10^{-3}$& $-220$& $100$& $15.25 \pm 0.05$& Y&  \\
S\sion{}& 1253& 1253.811& $1.09\times 10^{-2}$& $-220$& $100$& $15.35 \pm 0.05$& Y&  \\
S\sion{}& --& --& --& --& --& --& --& $15.33 \pm 0.05$\\
\hline
Fe\sion{}& 1608& 1608.451& $5.80\times 10^{-2}$& $-220$& $100$& $> 15.15$& Y&  \\
Fe\sion{}& 1611& 1611.200& $1.36\times 10^{-3}$& $-220$& $100$& $< 15.16$& Y&  \\
Fe\sion{}& --& --& --& --& --& --& --& $> 15.15$\\
\hline
Ni\sion{}& 1317& 1317.217& $5.71\times 10^{-2}$& $-220$& $100$& $14.06 \pm 0.07$& Y&  \\
Ni\sion{}& 1709& 1709.604& $3.24\times 10^{-2}$& $-220$& $100$& $13.91 \pm 0.12$& Y&  \\
Ni\sion{}& 1741& 1741.553& $4.27\times 10^{-2}$& $-220$& $100$& $14.04 \pm 0.07$& Y&  \\
Ni\sion{}& --& --& --& --& --& --& --& $14.02 \pm 0.05$\\
\hline
\end{tabular}
\end{center}
\end{table}

\begin{table}
\begin{center}
\caption{Metal column densities for J1142+0701 (\zabs{}=$1.84$)}
\label{tab:J11420701}
\begin{tabular}{lcccccccc}
\hline
Ion& Line& $\lambda$& $f$& $v_{\rm min}$& $v_{\rm max}$& logN(X)& Included& logN$_{\rm adopt}$\\
 &  & \AA{}&  & \kms{}& \kms{}&  &  &  \\
\hline
Si\sion{}& 1304& 1304.370& $9.40\times 10^{-2}$& $-220$& $230$& $> 15.21$& Y&  \\
Si\sion{}& 1526& 1526.707& $1.27\times 10^{-1}$& $-220$& $230$& $> 15.01$& Y&  \\
Si\sion{}& 1808& 1808.013& $2.19\times 10^{-3}$& $-220$& $230$& $16.15 \pm 0.13$& Y&  \\
Si\sion{}& --& --& --& --& --& --& --& $16.15 \pm 0.13$\\
\hline
S\sion{}& 1250& 1250.584& $5.43\times 10^{-3}$& $-200$& $40$& $> 15.69$& N&  \\
S\sion{}& 1253& 1253.811& $1.09\times 10^{-2}$& $-200$& $40$& $> 15.52$& N&  \\
\hline
\end{tabular}
\end{center}
\end{table}

\begin{table}
\begin{center}
\caption{Metal column densities for J1155+0530 (\zabs{}=$3.33$)}
\label{tab:J11550530}
\begin{tabular}{lcccccccc}
\hline
Ion& Line& $\lambda$& $f$& $v_{\rm min}$& $v_{\rm max}$& logN(X)& Included& logN$_{\rm adopt}$\\
 &  & \AA{}&  & \kms{}& \kms{}&  &  &  \\
\hline
Si\sion{}& 1304& 1304.370& $9.40\times 10^{-2}$& $-150$& $220$& $> 15.26$& N&  \\
Si\sion{}& 1526& 1526.707& $1.27\times 10^{-1}$& $-150$& $220$& $> 15.08$& N&  \\
Si\sion{}& 1808& 1808.013& $2.19\times 10^{-3}$& $-70$& $220$& $15.94 \pm 0.05$& Y&  \\
Si\sion{}& --& --& --& --& --& --& --& $15.94 \pm 0.05$\\
\hline
S\sion{}& 1250& 1250.584& $5.43\times 10^{-3}$& $-70$& $220$& $15.61 \pm 0.05$& Y&  \\
S\sion{}& 1253& 1253.811& $1.09\times 10^{-2}$& $-60$& $190$& $15.35 \pm 0.05$& Y&  \\
S\sion{}& --& --& --& --& --& --& --& $15.40 \pm 0.05$\\
\hline
Cr\sion{}& 2066& 2066.161& $5.15\times 10^{-2}$& $-70$& $220$& $13.36 \pm 0.09$& Y&  \\
Cr\sion{}& --& --& --& --& --& --& --& $13.36 \pm 0.09$\\
\hline
Fe\sion{}& 1608& 1608.451& $5.80\times 10^{-2}$& $-150$& $220$& $> 15.11$& N&  \\
Fe\sion{}& 1611& 1611.200& $1.36\times 10^{-3}$& $-70$& $160$& $15.37 \pm 0.05$& Y&  \\
Fe\sion{}& --& --& --& --& --& --& --& $15.37 \pm 0.05$\\
\hline
Ni\sion{}& 1317& 1317.217& $5.71\times 10^{-2}$& $-70$& $220$& $13.95 \pm 0.05$& Y&  \\
Ni\sion{}& 1370& 1370.131& $5.88\times 10^{-2}$& $-70$& $220$& $14.16 \pm 0.05$& Y&  \\
Ni\sion{}& 1454& 1454.842& $3.23\times 10^{-2}$& $-70$& $220$& $14.09 \pm 0.05$& Y&  \\
Ni\sion{}& 1709& 1709.604& $3.24\times 10^{-2}$& $-70$& $220$& $13.89 \pm 0.05$& Y&  \\
Ni\sion{}& 1741& 1741.553& $4.27\times 10^{-2}$& $-70$& $220$& $14.01 \pm 0.05$& Y&  \\
Ni\sion{}& 1751& 1751.916& $2.77\times 10^{-2}$& $-70$& $220$& $14.18 \pm 0.05$& Y&  \\
Ni\sion{}& --& --& --& --& --& --& --& $14.07 \pm 0.05$\\
\hline
Zn\sion{}& 2026& 2026.136& $4.89\times 10^{-1}$& $-70$& $220$& $12.89 \pm 0.07$& Y&  \\
Zn\sion{}& --& --& --& --& --& --& --& $12.89 \pm 0.07$\\
\hline
\end{tabular}
\end{center}
\end{table}

\begin{table}
\begin{center}
\caption{Metal column densities for J1305+0924 (\zabs{}=$2.02$)}
\label{tab:J13050924}
\begin{tabular}{lcccccccc}
\hline
Ion& Line& $\lambda$& $f$& $v_{\rm min}$& $v_{\rm max}$& logN(X)& Included& logN$_{\rm adopt}$\\
 &  & \AA{}&  & \kms{}& \kms{}&  &  &  \\
\hline
Si\sion{}& 1260& 1260.422& $1.01\times 10^{0}$& $-150$& $160$& $> 14.29$& N&  \\
Si\sion{}& 1304& 1304.370& $9.40\times 10^{-2}$& $-150$& $160$& $> 15.14$& N&  \\
Si\sion{}& 1808& 1808.013& $2.19\times 10^{-3}$& $-150$& $160$& $15.75 \pm 0.05$& Y&  \\
Si\sion{}& --& --& --& --& --& --& --& $15.75 \pm 0.05$\\
\hline
S\sion{}& 1250& 1250.584& $5.43\times 10^{-3}$& $-150$& $160$& $15.45 \pm 0.05$& Y&  \\
S\sion{}& 1253& 1253.811& $1.09\times 10^{-2}$& $-150$& $160$& $15.39 \pm 0.05$& Y&  \\
S\sion{}& 1259& 1259.519& $1.66\times 10^{-2}$& $-150$& $75$& $15.36 \pm 0.05$& Y&  \\
S\sion{}& --& --& --& --& --& --& --& $15.39 \pm 0.05$\\
\hline
Fe\sion{}& 1608& 1608.451& $5.80\times 10^{-2}$& $-150$& $160$& $> 15.10$& N&  \\
Fe\sion{}& 1611& 1611.200& $1.36\times 10^{-3}$& $-50$& $40$& $15.21 \pm 0.14$& Y&  \\
Fe\sion{}& --& --& --& --& --& --& --& $15.21 \pm 0.14$\\
\hline
Ni\sion{}& 1317& 1317.217& $5.71\times 10^{-2}$& $-150$& $160$& $14.31 \pm 0.05$& Y&  \\
Ni\sion{}& 1709& 1709.604& $3.24\times 10^{-2}$& $-150$& $160$& $14.57 \pm 0.05$& Y&  \\
Ni\sion{}& 1741& 1741.553& $4.27\times 10^{-2}$& $-150$& $160$& $14.28 \pm 0.06$& Y&  \\
Ni\sion{}& --& --& --& --& --& --& --& $14.36 \pm 0.05$\\
\hline
\end{tabular}
\end{center}
\end{table}

\clearpage

\begin{table}
\begin{center}
\caption{Metal column densities for J1310+5424 (\zabs{}=$1.80$)}
\label{tab:J13105424}
\begin{tabular}{lcccccccc}
\hline
Ion& Line& $\lambda$& $f$& $v_{\rm min}$& $v_{\rm max}$& logN(X)& Included& logN$_{\rm adopt}$\\
 &  & \AA{}&  & \kms{}& \kms{}&  &  &  \\
\hline
Si\sion{}& 1260& 1260.422& $1.01\times 10^{0}$& $-110$& $60$& $> 14.10$& N&  \\
Si\sion{}& 1304& 1304.370& $9.40\times 10^{-2}$& $-100$& $100$& $> 15.13$& Y&  \\
Si\sion{}& 1526& 1526.707& $1.27\times 10^{-1}$& $-100$& $100$& $> 14.95$& Y&  \\
Si\sion{}& 1808& 1808.013& $2.19\times 10^{-3}$& $-100$& $100$& $16.44 \pm 0.05$& Y&  \\
Si\sion{}& --& --& --& --& --& --& --& $16.44 \pm 0.05$\\
\hline
S\sion{}& 1250& 1250.584& $5.43\times 10^{-3}$& $-80$& $90$& $> 16.05$& Y&  \\
S\sion{}& 1253& 1253.811& $1.09\times 10^{-2}$& $-80$& $60$& $> 15.84$& Y&  \\
S\sion{}& --& --& --& --& --& --& --& $> 16.05$\\
\hline
Cr\sion{}& 2066& 2066.161& $5.15\times 10^{-2}$& $-60$& $80$& $13.99 \pm 0.05$& Y&  \\
Cr\sion{}& --& --& --& --& --& --& --& $13.99 \pm 0.05$\\
\hline
Fe\sion{}& 1608& 1608.451& $5.80\times 10^{-2}$& $-100$& $80$& $> 15.10$& Y&  \\
Fe\sion{}& 1611& 1611.200& $1.36\times 10^{-3}$& $-60$& $90$& $15.64 \pm 0.05$& Y&  \\
Fe\sion{}& --& --& --& --& --& --& --& $15.64 \pm 0.05$\\
\hline
Ni\sion{}& 1317& 1317.217& $5.71\times 10^{-2}$& $-60$& $90$& $14.44 \pm 0.05$& Y&  \\
Ni\sion{}& 1454& 1454.842& $3.23\times 10^{-2}$& $-60$& $100$& $14.43 \pm 0.05$& Y&  \\
Ni\sion{}& 1703& 1703.405& $6.00\times 10^{-3}$& $-60$& $90$& $14.48 \pm 0.07$& Y&  \\
Ni\sion{}& 1709& 1709.604& $3.24\times 10^{-2}$& $-60$& $90$& $14.44 \pm 0.05$& Y&  \\
Ni\sion{}& 1741& 1741.553& $4.27\times 10^{-2}$& $-60$& $90$& $14.48 \pm 0.05$& Y&  \\
Ni\sion{}& 1751& 1751.916& $2.77\times 10^{-2}$& $-50$& $90$& $14.47 \pm 0.05$& Y&  \\
Ni\sion{}& --& --& --& --& --& --& --& $14.45 \pm 0.05$\\
\hline
Zn\sion{}& 2026& 2026.136& $4.89\times 10^{-1}$& $-60$& $100$& $13.57 \pm 0.05$& Y&  \\
Zn\sion{}& --& --& --& --& --& --& --& $13.57 \pm 0.05$\\
\hline
\end{tabular}
\end{center}
\end{table}

\begin{table}
\begin{center}
\caption{Metal column densities for J1313+1441 (\zabs{}=$1.79$)}
\label{tab:J13131441}
\begin{tabular}{lcccccccc}
\hline
Ion& Line& $\lambda$& $f$& $v_{\rm min}$& $v_{\rm max}$& logN(X)& Included& logN$_{\rm adopt}$\\
 &  & \AA{}&  & \kms{}& \kms{}&  &  &  \\
\hline
Si\sion{}& 1304& 1304.370& $9.40\times 10^{-2}$& $-160$& $195$& $> 15.29$& Y&  \\
Si\sion{}& 1526& 1526.707& $1.27\times 10^{-1}$& $-160$& $195$& $> 15.10$& Y&  \\
Si\sion{}& 1808& 1808.013& $2.19\times 10^{-3}$& $-160$& $195$& $16.12 \pm 0.05$& Y&  \\
Si\sion{}& --& --& --& --& --& --& --& $16.12 \pm 0.05$\\
\hline
S\sion{}& 1250& 1250.584& $5.43\times 10^{-3}$& $-160$& $195$& $> 15.75$& Y&  \\
S\sion{}& 1253& 1253.811& $1.09\times 10^{-2}$& $-160$& $195$& $> 15.85$& N&  \\
S\sion{}& --& --& --& --& --& --& --& $> 15.75$\\
\hline
Cr\sion{}& 2056& 2056.254& $1.05\times 10^{-1}$& $-160$& $195$& $13.60 \pm 0.05$& Y&  \\
Cr\sion{}& 2066& 2066.161& $5.15\times 10^{-2}$& $-160$& $195$& $13.65 \pm 0.06$& Y&  \\
Cr\sion{}& --& --& --& --& --& --& --& $13.61 \pm 0.05$\\
\hline
Fe\sion{}& 1608& 1608.451& $5.80\times 10^{-2}$& $-160$& $195$& $> 15.20$& N&  \\
Fe\sion{}& 1611& 1611.200& $1.36\times 10^{-3}$& $-160$& $195$& $15.55 \pm 0.05$& Y&  \\
Fe\sion{}& --& --& --& --& --& --& --& $15.55 \pm 0.05$\\
\hline
Ni\sion{}& 1317& 1317.217& $5.71\times 10^{-2}$& $-160$& $195$& $14.30 \pm 0.05$& Y&  \\
Ni\sion{}& 1370& 1370.131& $5.88\times 10^{-2}$& $-100$& $195$& $14.26 \pm 0.05$& Y&  \\
Ni\sion{}& 1454& 1454.842& $3.23\times 10^{-2}$& $-160$& $195$& $14.09 \pm 0.05$& Y&  \\
Ni\sion{}& 1709& 1709.604& $3.24\times 10^{-2}$& $-160$& $195$& $14.26 \pm 0.05$& Y&  \\
Ni\sion{}& 1741& 1741.553& $4.27\times 10^{-2}$& $-160$& $195$& $14.29 \pm 0.05$& Y&  \\
Ni\sion{}& 1751& 1751.916& $2.77\times 10^{-2}$& $-160$& $195$& $14.37 \pm 0.05$& Y&  \\
Ni\sion{}& --& --& --& --& --& --& --& $14.27 \pm 0.05$\\
\hline
Zn\sion{}& 2026& 2026.136& $4.89\times 10^{-1}$& $-160$& $195$& $13.30 \pm 0.05$& Y&  \\
Zn\sion{}& --& --& --& --& --& --& --& $13.30 \pm 0.05$\\
\hline
\end{tabular}
\end{center}
\end{table}

\begin{table}
\begin{center}
\caption{Metal column densities for J1335+0824 (\zabs{}=$1.86$)}
\label{tab:J13350824}
\begin{tabular}{lcccccccc}
\hline
Ion& Line& $\lambda$& $f$& $v_{\rm min}$& $v_{\rm max}$& logN(X)& Included& logN$_{\rm adopt}$\\
 &  & \AA{}&  & \kms{}& \kms{}&  &  &  \\
\hline
Si\sion{}& 1304& 1304.370& $9.40\times 10^{-2}$& $-170$& $170$& $> 15.23$& N&  \\
Si\sion{}& 1526& 1526.707& $1.27\times 10^{-1}$& $-170$& $170$& $> 15.05$& N&  \\
Si\sion{}& 1808& 1808.013& $2.19\times 10^{-3}$& $-120$& $130$& $15.73 \pm 0.05$& Y&  \\
Si\sion{}& --& --& --& --& --& --& --& $15.73 \pm 0.05$\\
\hline
S\sion{}& 1250& 1250.584& $5.43\times 10^{-3}$& $-120$& $130$& $15.26 \pm 0.05$& Y&  \\
S\sion{}& 1253& 1253.811& $1.09\times 10^{-2}$& $-120$& $130$& $15.30 \pm 0.05$& Y&  \\
S\sion{}& --& --& --& --& --& --& --& $15.29 \pm 0.05$\\
\hline
Mn\sion{}& 1197& 1197.184& $1.57\times 10^{-1}$& $-120$& $130$& $13.70 \pm 0.10$& Y&  \\
Mn\sion{}& --& --& --& --& --& --& --& $13.70 \pm 0.10$\\
\hline
Cr\sion{}& 2056& 2056.254& $1.05\times 10^{-1}$& $-120$& $75$& $13.81 \pm 0.05$& Y&  \\
Cr\sion{}& --& --& --& --& --& --& --& $13.81 \pm 0.05$\\
\hline
Fe\sion{}& 1608& 1608.451& $5.80\times 10^{-2}$& $-170$& $170$& $> 15.17$& Y&  \\
Fe\sion{}& --& --& --& --& --& --& --& $> 15.17$\\
\hline
Ni\sion{}& 1317& 1317.217& $5.71\times 10^{-2}$& $-120$& $130$& $> 14.26$& Y&  \\
Ni\sion{}& 1741& 1741.553& $4.27\times 10^{-2}$& $-120$& $130$& $14.29 \pm 0.05$& Y&  \\
Ni\sion{}& --& --& --& --& --& --& --& $14.29 \pm 0.05$\\
\hline
\end{tabular}
\end{center}
\end{table}

\begin{table}
\begin{center}
\caption{Metal column densities for J1417+4132 (\zabs{}=$1.95$)}
\label{tab:J14174132}
\begin{tabular}{lcccccccc}
\hline
Ion& Line& $\lambda$& $f$& $v_{\rm min}$& $v_{\rm max}$& logN(X)& Included& logN$_{\rm adopt}$\\
 &  & \AA{}&  & \kms{}& \kms{}&  &  &  \\
\hline
Si\sion{}& 1304& 1304.370& $9.40\times 10^{-2}$& $-105$& $390$& $> 15.16$& Y&  \\
Si\sion{}& 1526& 1526.707& $1.27\times 10^{-1}$& $-105$& $390$& $> 14.96$& Y&  \\
Si\sion{}& 1808& 1808.013& $2.19\times 10^{-3}$& $-105$& $390$& $> 16.42$& Y&  \\
Si\sion{}& --& --& --& --& --& --& --& $> 16.42$\\
\hline
Cr\sion{}& 2056& 2056.254& $1.05\times 10^{-1}$& $-105$& $390$& $14.03 \pm 0.05$& Y&  \\
Cr\sion{}& 2066& 2066.161& $5.15\times 10^{-2}$& $-45$& $390$& $14.08 \pm 0.05$& Y&  \\
Cr\sion{}& --& --& --& --& --& --& --& $14.04 \pm 0.05$\\
\hline
Fe\sion{}& 1608& 1608.451& $5.80\times 10^{-2}$& $-105$& $390$& $> 15.22$& N&  \\
Fe\sion{}& 1611& 1611.200& $1.36\times 10^{-3}$& $-105$& $390$& $15.58 \pm 0.05$& Y&  \\
Fe\sion{}& --& --& --& --& --& --& --& $15.58 \pm 0.05$\\
\hline
Ni\sion{}& 1317& 1317.217& $5.71\times 10^{-2}$& $-105$& $390$& $14.62 \pm 0.05$& Y&  \\
Ni\sion{}& 1370& 1370.131& $5.88\times 10^{-2}$& $-105$& $390$& $14.58 \pm 0.05$& Y&  \\
Ni\sion{}& 1454& 1454.842& $3.23\times 10^{-2}$& $-105$& $150$& $14.46 \pm 0.05$& Y&  \\
Ni\sion{}& 1467a& 1467.259& $6.30\times 10^{-3}$& $-30$& $50$& $13.92 \pm 0.11$& N&  \\
Ni\sion{}& 1703& 1703.405& $6.00\times 10^{-3}$& $-105$& $100$& $14.36 \pm 0.05$& Y&  \\
Ni\sion{}& 1709& 1709.604& $3.24\times 10^{-2}$& $-105$& $250$& $14.56 \pm 0.05$& Y&  \\
Ni\sion{}& 1741& 1741.553& $4.27\times 10^{-2}$& $-105$& $390$& $14.52 \pm 0.05$& Y&  \\
Ni\sion{}& 1751& 1751.916& $2.77\times 10^{-2}$& $-105$& $390$& $14.53 \pm 0.05$& Y&  \\
Ni\sion{}& --& --& --& --& --& --& --& $14.55 \pm 0.05$\\
\hline
Zn\sion{}& 2026& 2026.136& $4.89\times 10^{-1}$& $-105$& $390$& $13.55 \pm 0.05$& Y&  \\
Zn\sion{}& --& --& --& --& --& --& --& $13.55 \pm 0.05$\\
\hline
\end{tabular}
\end{center}
\end{table}

\begin{table}
\begin{center}
\caption{Metal column densities for J1454+0941 (\zabs{}=$1.79$)}
\label{tab:J14540941}
\begin{tabular}{lcccccccc}
\hline
Ion& Line& $\lambda$& $f$& $v_{\rm min}$& $v_{\rm max}$& logN(X)& Included& logN$_{\rm adopt}$\\
 &  & \AA{}&  & \kms{}& \kms{}&  &  &  \\
\hline
Si\sion{}& 1304& 1304.370& $9.40\times 10^{-2}$& $-180$& $230$& $> 15.10$& N&  \\
Si\sion{}& 1526& 1526.707& $1.27\times 10^{-1}$& $-180$& $230$& $> 14.98$& N&  \\
Si\sion{}& 1808& 1808.013& $2.19\times 10^{-3}$& $-80$& $70$& $15.47 \pm 0.05$& Y&  \\
Si\sion{}& --& --& --& --& --& --& --& $15.47 \pm 0.05$\\
\hline
S\sion{}& 1250& 1250.584& $5.43\times 10^{-3}$& $-80$& $70$& $15.25 \pm 0.06$& Y&  \\
S\sion{}& 1253& 1253.811& $1.09\times 10^{-2}$& $-80$& $70$& $> 15.35$& Y&  \\
S\sion{}& --& --& --& --& --& --& --& $15.25 \pm 0.06$\\
\hline
Cr\sion{}& 2056& 2056.254& $1.05\times 10^{-1}$& $-80$& $70$& $13.30 \pm 0.09$& Y&  \\
Cr\sion{}& --& --& --& --& --& --& --& $13.30 \pm 0.09$\\
\hline
Fe\sion{}& 1143& 1143.226& $1.77\times 10^{-2}$& $-80$& $70$& $> 15.26$& Y&  \\
Fe\sion{}& 1608& 1608.451& $5.80\times 10^{-2}$& $-180$& $230$& $> 14.99$& N&  \\
Fe\sion{}& 1611& 1611.200& $1.36\times 10^{-3}$& $-80$& $70$& $15.02 \pm 0.12$& Y&  \\
Fe\sion{}& --& --& --& --& --& --& --& $15.02 \pm 0.12$\\
\hline
Ni\sion{}& 1454& 1454.842& $3.23\times 10^{-2}$& $-80$& $70$& $13.86 \pm 0.14$& Y&  \\
Ni\sion{}& 1709& 1709.604& $3.24\times 10^{-2}$& $-80$& $70$& $13.80 \pm 0.08$& Y&  \\
Ni\sion{}& 1741& 1741.553& $4.27\times 10^{-2}$& $-80$& $70$& $13.85 \pm 0.05$& Y&  \\
Ni\sion{}& 1751& 1751.916& $2.77\times 10^{-2}$& $-80$& $70$& $13.90 \pm 0.08$& Y&  \\
Ni\sion{}& --& --& --& --& --& --& --& $13.85 \pm 0.05$\\
\hline
Zn\sion{}& 2026& 2026.136& $4.89\times 10^{-1}$& $-80$& $70$& $12.68 \pm 0.06$& Y&  \\
Zn\sion{}& 2062& 2062.664& $2.56\times 10^{-1}$& $-80$& $70$& $12.87 \pm 0.09$& Y&  \\
Zn\sion{}& --& --& --& --& --& --& --& $12.72 \pm 0.05$\\
\hline
\end{tabular}
\end{center}
\end{table}

\begin{table}
\begin{center}
\caption{Metal column densities for J1509+1113 (\zabs{}=$2.03$)}
\label{tab:J15091113}
\begin{tabular}{lcccccccc}
\hline
Ion& Line& $\lambda$& $f$& $v_{\rm min}$& $v_{\rm max}$& logN(X)& Included& logN$_{\rm adopt}$\\
 &  & \AA{}&  & \kms{}& \kms{}&  &  &  \\
\hline
Si\sion{}& 1304& 1304.370& $9.40\times 10^{-2}$& $-260$& $160$& $> 15.14$& Y&  \\
Si\sion{}& 1526& 1526.707& $1.27\times 10^{-1}$& $-260$& $160$& $> 14.94$& Y&  \\
Si\sion{}& 1808& 1808.013& $2.19\times 10^{-3}$& $-260$& $160$& $16.04 \pm 0.05$& Y&  \\
Si\sion{}& --& --& --& --& --& --& --& $16.04 \pm 0.05$\\
\hline
S\sion{}& 1250& 1250.584& $5.43\times 10^{-3}$& $-150$& $70$& $15.81 \pm 0.05$& Y&  \\
S\sion{}& 1253& 1253.811& $1.09\times 10^{-2}$& $-150$& $70$& $15.60 \pm 0.05$& Y&  \\
S\sion{}& --& --& --& --& --& --& --& $15.69 \pm 0.05$\\
\hline
Fe\sion{}& 1608& 1608.451& $5.80\times 10^{-2}$& $-120$& $160$& $> 15.16$& N&  \\
Fe\sion{}& 1611& 1611.200& $1.36\times 10^{-3}$& $-150$& $70$& $15.48 \pm 0.07$& Y&  \\
Fe\sion{}& --& --& --& --& --& --& --& $15.48 \pm 0.07$\\
\hline
Ni\sion{}& 1317& 1317.217& $5.71\times 10^{-2}$& $-150$& $70$& $14.27 \pm 0.05$& Y&  \\
Ni\sion{}& 1709& 1709.604& $3.24\times 10^{-2}$& $-150$& $70$& $14.46 \pm 0.06$& Y&  \\
Ni\sion{}& 1741& 1741.553& $4.27\times 10^{-2}$& $-150$& $70$& $14.41 \pm 0.05$& Y&  \\
Ni\sion{}& 1751& 1751.916& $2.77\times 10^{-2}$& $-150$& $70$& $14.55 \pm 0.05$& Y&  \\
Ni\sion{}& --& --& --& --& --& --& --& $14.41 \pm 0.05$\\
\hline
\end{tabular}
\end{center}
\end{table}

\begin{table}
\begin{center}
\caption{Metal column densities for J1524+1030 (\zabs{}=$1.94$)}
\label{tab:J15241030}
\begin{tabular}{lcccccccc}
\hline
Ion& Line& $\lambda$& $f$& $v_{\rm min}$& $v_{\rm max}$& logN(X)& Included& logN$_{\rm adopt}$\\
 &  & \AA{}&  & \kms{}& \kms{}&  &  &  \\
\hline
Si\sion{}& 1526& 1526.707& $1.27\times 10^{-1}$& $-370$& $70$& $> 15.26$& Y&  \\
Si\sion{}& 1808& 1808.013& $2.19\times 10^{-3}$& $-370$& $70$& $> 16.24$& Y&  \\
Si\sion{}& --& --& --& --& --& --& --& $> 16.24$\\
\hline
S\sion{}& 1253& 1253.811& $1.09\times 10^{-2}$& $-60$& $60$& $> 15.63$& Y&  \\
S\sion{}& --& --& --& --& --& --& --& $> 15.63$\\
\hline
Cr\sion{}& 2056& 2056.254& $1.05\times 10^{-1}$& $-50$& $70$& $13.57 \pm 0.05$& Y&  \\
Cr\sion{}& --& --& --& --& --& --& --& $13.57 \pm 0.05$\\
\hline
Fe\sion{}& 1608& 1608.451& $5.80\times 10^{-2}$& $-370$& $70$& $> 15.34$& N&  \\
Fe\sion{}& 1611& 1611.200& $1.36\times 10^{-3}$& $-100$& $70$& $15.44 \pm 0.05$& Y&  \\
Fe\sion{}& --& --& --& --& --& --& --& $15.44 \pm 0.05$\\
\hline
Ni\sion{}& 1370& 1370.131& $5.88\times 10^{-2}$& $-370$& $70$& $> 14.47$& Y&  \\
Ni\sion{}& 1709& 1709.604& $3.24\times 10^{-2}$& $-370$& $70$& $14.53 \pm 0.05$& Y&  \\
Ni\sion{}& 1741& 1741.553& $4.27\times 10^{-2}$& $-100$& $70$& $> 14.31$& Y&  \\
Ni\sion{}& --& --& --& --& --& --& --& $14.53 \pm 0.05$\\
\hline
Zn\sion{}& 2026& 2026.136& $4.89\times 10^{-1}$& $-70$& $70$& $> 13.47$& Y&  \\
Zn\sion{}& 2062& 2062.664& $2.56\times 10^{-1}$& $-70$& $70$& $> 13.52$& Y&  \\
Zn\sion{}& --& --& --& --& --& --& --& $> 13.52$\\
\hline
\end{tabular}
\end{center}
\end{table}

\begin{table}
\begin{center}
\caption{Metal column densities for J1552+4910 (\zabs{}=$1.96$)}
\label{tab:J15524910}
\begin{tabular}{lcccccccc}
\hline
Ion& Line& $\lambda$& $f$& $v_{\rm min}$& $v_{\rm max}$& logN(X)& Included& logN$_{\rm adopt}$\\
 &  & \AA{}&  & \kms{}& \kms{}&  &  &  \\
\hline
Si\sion{}& 1526& 1526.707& $1.27\times 10^{-1}$& $-135$& $120$& $> 14.96$& N&  \\
Si\sion{}& 1808& 1808.013& $2.19\times 10^{-3}$& $-135$& $120$& $15.98 \pm 0.05$& Y&  \\
Si\sion{}& --& --& --& --& --& --& --& $15.98 \pm 0.05$\\
\hline
S\sion{}& 1250& 1250.584& $5.43\times 10^{-3}$& $-135$& $120$& $15.37 \pm 0.05$& Y&  \\
S\sion{}& 1259& 1259.519& $1.66\times 10^{-2}$& $-135$& $70$& $15.34 \pm 0.05$& Y&  \\
S\sion{}& --& --& --& --& --& --& --& $15.34 \pm 0.05$\\
\hline
\end{tabular}
\end{center}
\end{table}

\begin{table}
\begin{center}
\caption{Metal column densities for J1555+4800 (\zabs{}=$2.39$)}
\label{tab:J15554800}
\begin{tabular}{lcccccccc}
\hline
Ion& Line& $\lambda$& $f$& $v_{\rm min}$& $v_{\rm max}$& logN(X)& Included& logN$_{\rm adopt}$\\
 &  & \AA{}&  & \kms{}& \kms{}&  &  &  \\
\hline
Si\sion{}& 1304& 1304.370& $9.40\times 10^{-2}$& $-265$& $320$& $> 15.46$& Y&  \\
Si\sion{}& 1526& 1526.707& $1.27\times 10^{-1}$& $-265$& $320$& $> 15.33$& Y&  \\
Si\sion{}& 1808& 1808.013& $2.19\times 10^{-3}$& $-265$& $320$& $16.55 \pm 0.05$& Y&  \\
Si\sion{}& --& --& --& --& --& --& --& $16.55 \pm 0.05$\\
\hline
Cr\sion{}& 2056& 2056.254& $1.05\times 10^{-1}$& $-265$& $320$& $14.23 \pm 0.05$& Y&  \\
Cr\sion{}& 2066& 2066.161& $5.15\times 10^{-2}$& $-265$& $320$& $14.06 \pm 0.05$& Y&  \\
Cr\sion{}& --& --& --& --& --& --& --& $14.19 \pm 0.05$\\
\hline
Fe\sion{}& 1608& 1608.451& $5.80\times 10^{-2}$& $-265$& $320$& $> 15.45$& N&  \\
Fe\sion{}& 1611& 1611.200& $1.36\times 10^{-3}$& $-265$& $320$& $15.84 \pm 0.05$& Y&  \\
Fe\sion{}& --& --& --& --& --& --& --& $15.84 \pm 0.05$\\
\hline
Ni\sion{}& 1709& 1709.604& $3.24\times 10^{-2}$& $-265$& $320$& $14.78 \pm 0.05$& Y&  \\
Ni\sion{}& 1741& 1741.553& $4.27\times 10^{-2}$& $-265$& $320$& $14.78 \pm 0.05$& Y&  \\
Ni\sion{}& --& --& --& --& --& --& --& $14.78 \pm 0.05$\\
\hline
Zn\sion{}& 2026& 2026.136& $4.89\times 10^{-1}$& $-265$& $70$& $< 13.95$& Y&  \\
Zn\sion{}& --& --& --& --& --& --& --& $< 13.95$\\
\hline
\end{tabular}
\end{center}
\end{table}

\begin{table}
\begin{center}
\caption{Metal column densities for J1610+4724 (\zabs{}=$2.51$)}
\label{tab:J16104724}
\begin{tabular}{lcccccccc}
\hline
Ion& Line& $\lambda$& $f$& $v_{\rm min}$& $v_{\rm max}$& logN(X)& Included& logN$_{\rm adopt}$\\
 &  & \AA{}&  & \kms{}& \kms{}&  &  &  \\
\hline
Si\sion{}& 1304& 1304.370& $9.40\times 10^{-2}$& $-180$& $380$& $> 15.42$& Y&  \\
Si\sion{}& 1526& 1526.707& $1.27\times 10^{-1}$& $-145$& $380$& $> 15.22$& Y&  \\
Si\sion{}& 1808& 1808.013& $2.19\times 10^{-3}$& $-180$& $380$& $16.16 \pm 0.05$& Y&  \\
Si\sion{}& --& --& --& --& --& --& --& $16.16 \pm 0.05$\\
\hline
Cr\sion{}& 2056& 2056.254& $1.05\times 10^{-1}$& $-70$& $180$& $13.91 \pm 0.05$& Y&  \\
Cr\sion{}& --& --& --& --& --& --& --& $13.91 \pm 0.05$\\
\hline
Fe\sion{}& 1608& 1608.451& $5.80\times 10^{-2}$& $-180$& $320$& $> 15.33$& N&  \\
Fe\sion{}& 1611& 1611.200& $1.36\times 10^{-3}$& $-120$& $140$& $15.78 \pm 0.05$& Y&  \\
Fe\sion{}& 2249& 2249.877& $1.82\times 10^{-3}$& $-70$& $120$& $15.59 \pm 0.05$& Y&  \\
Fe\sion{}& 2260& 2260.780& $2.44\times 10^{-3}$& $-70$& $120$& $15.61 \pm 0.05$& Y&  \\
Fe\sion{}& --& --& --& --& --& --& --& $15.62 \pm 0.05$\\
\hline
Ni\sion{}& 1703& 1703.405& $6.00\times 10^{-3}$& $-50$& $140$& $14.48 \pm 0.09$& Y&  \\
Ni\sion{}& 1709& 1709.604& $3.24\times 10^{-2}$& $-50$& $140$& $14.61 \pm 0.05$& Y&  \\
Ni\sion{}& 1741& 1741.553& $4.27\times 10^{-2}$& $-50$& $140$& $14.54 \pm 0.05$& Y&  \\
Ni\sion{}& 1751& 1751.916& $2.77\times 10^{-2}$& $-50$& $140$& $14.58 \pm 0.05$& Y&  \\
Ni\sion{}& --& --& --& --& --& --& --& $14.58 \pm 0.05$\\
\hline
Zn\sion{}& 2026& 2026.136& $4.89\times 10^{-1}$& $-70$& $180$& $13.56 \pm 0.05$& Y&  \\
Zn\sion{}& --& --& --& --& --& --& --& $13.56 \pm 0.05$\\
\hline
\end{tabular}
\end{center}
\end{table}

\begin{table}
\begin{center}
\caption{Metal column densities for J1629+0913 (\zabs{}=$1.90$)}
\label{tab:J16290913}
\begin{tabular}{lcccccccc}
\hline
Ion& Line& $\lambda$& $f$& $v_{\rm min}$& $v_{\rm max}$& logN(X)& Included& logN$_{\rm adopt}$\\
 &  & \AA{}&  & \kms{}& \kms{}&  &  &  \\
\hline
Si\sion{}& 1304& 1304.370& $9.40\times 10^{-2}$& $-170$& $400$& $> 15.38$& N&  \\
Si\sion{}& 1526& 1526.707& $1.27\times 10^{-1}$& $-170$& $420$& $> 15.15$& N&  \\
Si\sion{}& 1808& 1808.013& $2.19\times 10^{-3}$& $-170$& $50$& $15.32 \pm 0.06$& Y&  \\
Si\sion{}& --& --& --& --& --& --& --& $15.32 \pm 0.06$\\
\hline
S\sion{}& 1250& 1250.584& $5.43\times 10^{-3}$& $-170$& $50$& $15.09 \pm 0.05$& Y&  \\
S\sion{}& 1253& 1253.811& $1.09\times 10^{-2}$& $-170$& $50$& $15.29 \pm 0.05$& Y&  \\
S\sion{}& --& --& --& --& --& --& --& $15.24 \pm 0.05$\\
\hline
Cr\sion{}& 2056& 2056.254& $1.05\times 10^{-1}$& $-170$& $50$& $< 13.21$& Y&  \\
Cr\sion{}& --& --& --& --& --& --& --& $< 13.21$\\
\hline
Fe\sion{}& 1608& 1608.451& $5.80\times 10^{-2}$& $-170$& $140$& $> 14.93$& Y&  \\
Fe\sion{}& 1611& 1611.200& $1.36\times 10^{-3}$& $-170$& $50$& $< 15.07$& Y&  \\
Fe\sion{}& --& --& --& --& --& --& --& $> 14.93$\\
\hline
Ni\sion{}& 1317& 1317.217& $5.71\times 10^{-2}$& $-60$& $50$& $13.75 \pm 0.13$& Y&  \\
Ni\sion{}& 1370& 1370.131& $5.88\times 10^{-2}$& $-170$& $50$& $> 13.65$& Y&  \\
Ni\sion{}& 1741& 1741.553& $4.27\times 10^{-2}$& $-170$& $50$& $< 13.98$& Y&  \\
Ni\sion{}& --& --& --& --& --& --& --& $13.75 \pm 0.13$\\
\hline
Zn\sion{}& 2026& 2026.136& $4.89\times 10^{-1}$& $-30$& $50$& $12.68 \pm 0.08$& Y&  \\
Zn\sion{}& --& --& --& --& --& --& --& $12.68 \pm 0.08$\\
\hline
\end{tabular}
\end{center}
\end{table}

\begin{table}
\begin{center}
\caption{Metal column densities for Q1755+578 (\zabs{}=$1.97$)}
\label{tab:Q1755578}
\begin{tabular}{lcccccccc}
\hline
Ion& Line& $\lambda$& $f$& $v_{\rm min}$& $v_{\rm max}$& logN(X)& Included& logN$_{\rm adopt}$\\
 &  & \AA{}&  & \kms{}& \kms{}&  &  &  \\
\hline
Si\sion{}& 1304& 1304.370& $9.40\times 10^{-2}$& $-30$& $500$& $> 15.57$& Y&  \\
Si\sion{}& 1526& 1526.707& $1.27\times 10^{-1}$& $-200$& $500$& $> 15.43$& Y&  \\
Si\sion{}& 1808& 1808.013& $2.19\times 10^{-3}$& $-200$& $500$& $16.58 \pm 0.05$& Y&  \\
Si\sion{}& --& --& --& --& --& --& --& $16.58 \pm 0.05$\\
\hline
S\sion{}& 1253& 1253.811& $1.09\times 10^{-2}$& $-50$& $500$& $> 16.12$& Y&  \\
S\sion{}& --& --& --& --& --& --& --& $> 16.12$\\
\hline
\end{tabular}
\end{center}
\end{table}

\begin{table}
\begin{center}
\caption{Metal column densities for J2241+1225 (\zabs{}=$2.42$)}
\label{tab:J22411225}
\begin{tabular}{lcccccccc}
\hline
Ion& Line& $\lambda$& $f$& $v_{\rm min}$& $v_{\rm max}$& logN(X)& Included& logN$_{\rm adopt}$\\
 &  & \AA{}&  & \kms{}& \kms{}&  &  &  \\
\hline
Si\sion{}& 1304& 1304.370& $9.40\times 10^{-2}$& $-100$& $25$& $> 14.67$& Y&  \\
Si\sion{}& 1526& 1526.707& $1.27\times 10^{-1}$& $-90$& $25$& $> 14.49$& Y&  \\
Si\sion{}& --& --& --& --& --& --& --& $> 14.67$\\
\hline
S\sion{}& 1253& 1253.811& $1.09\times 10^{-2}$& $-100$& $25$& $> 15.01$& Y&  \\
S\sion{}& 1259& 1259.519& $1.66\times 10^{-2}$& $-100$& $25$& $> 14.91$& Y&  \\
S\sion{}& --& --& --& --& --& --& --& $> 15.01$\\
\hline
Mn\sion{}& 1197& 1197.184& $1.57\times 10^{-1}$& $-100$& $25$& $13.36 \pm 0.13$& Y&  \\
Mn\sion{}& --& --& --& --& --& --& --& $13.36 \pm 0.13$\\
\hline
Fe\sion{}& 1081& 1081.875& $1.40\times 10^{-2}$& $-100$& $25$& $> 15.02$& Y&  \\
Fe\sion{}& 1142& 1142.366& $4.20\times 10^{-3}$& $-100$& $25$& $15.01 \pm 0.08$& Y&  \\
Fe\sion{}& 1608& 1608.451& $5.80\times 10^{-2}$& $-100$& $25$& $> 14.45$& N&  \\
Fe\sion{}& 1611& 1611.200& $1.36\times 10^{-3}$& $-100$& $25$& $< 15.12$& Y&  \\
Fe\sion{}& --& --& --& --& --& --& --& $15.02 \pm 0.08$\\
\hline
Ni\sion{}& 1317& 1317.217& $5.71\times 10^{-2}$& $-100$& $25$& $13.89 \pm 0.05$& Y&  \\
Ni\sion{}& 1370& 1370.131& $5.88\times 10^{-2}$& $-100$& $25$& $13.65 \pm 0.10$& Y&  \\
Ni\sion{}& --& --& --& --& --& --& --& $13.83 \pm 0.05$\\
\hline
\end{tabular}
\end{center}
\end{table}

\end{document}